\documentclass[11pt]{amsart}
\usepackage{amssymb, amsfonts, stmaryrd,fancyhdr,amsmath,amsthm,amsbsy,amsfonts, titletoc} 
\usepackage{graphicx,rotating,wrapfig}   

\usepackage[utf8]{inputenc}
\usepackage{multirow,bigstrut}
\usepackage[usenames, dvipsnames]{color}
\usepackage[citecolor=blue]{hyperref}
\hypersetup{colorlinks=true,linkcolor=blue,urlcolor=blue}
\graphicspath{{Images/}}
\theoremstyle{plain}

\newcommand{\beq}{\begin{equation}}
\newcommand{\eeq}{\end{equation}}
\newcommand{\bna}{\begin{eqnarray}}
\newcommand{\ena}{\end{eqnarray}}
\newcommand{\bea}{\begin{eqnarray*}}
\newcommand{\eea}{\end{eqnarray*}}

\newcommand{\bt}[1]{ \begin{tabular} { #1 } }
\newcommand{\et} {\end{tabular}}
\newcommand{\normmm}[1]{{\left\vert\kern-0.25ex\left\vert\kern-0.25ex\left\vert #1 
    \right\vert\kern-0.25ex\right\vert\kern-0.25ex\right\vert}}

\def\bphi{m}

\def\epsA{\varrho_A}

\def\La{\ell_A}

\def\mrG{\mathrm G}

\def\cD{\mathcal{D}}
\def\cG{\mathcal{G}}
\def\mrI{\pi}

\def\phiro{\phi^{\rm 0}}
\def\psiro{\psi^{\rm 0}}

\def\cF{{\mathcal F}}

\def\bpm{\begin{pmatrix}}
\def\epm{\end{pmatrix}}

 \def\eDF{\mathrm{pDF}}
 \def\GDF{\mathrm{GDF}}

 \numberwithin{equation}{section}
 
\begin{document}
\title{\textbf{Parameterized Density Functional Models of Block Copolymer Melts} }
\author[Wang]{Sulin Wang}
\address{Sulin Wang, 
School of Mathematics and Hunan Provincial Key Laboratory of Intelligent Information Processing and Applied Mathematics, Hunan University, Changsha 410082, The P.R. of China, sulinw@hnu.edu.cn}

\author[Chen]{Yuan Chen}
\address{Yuan Chen, School of Science and Engineering, Chinese University of Hong Kong (Shenzhen), Shenzhen, Guangdong, The P.R. of China, chenyuan@cuhk.edu.cn}
\author[Tan]{Zengqiang Tan}
\address{Zengqiang Tan, School of Mathematical Sciences, Peking University, Beijing 100871, The P.R. of China, tzengqiang@163.com}
\author[Promislow]{Keith Promislow}
\address{Keith Promislow, Department of Mathematics, Michigan State University,
East Lansing, MI 48824, USA, promislo@msu.edu}
\date{\today}
\maketitle

\begin{abstract}
The derivation of density functional energies from the random phase approximation of self-consistent mean field theory is generalized and applied to a binary blend of diblock copolymers and homopolymers. A nonlocal transformation is incorporated into the density functional model prior to the strong segregation extrapolation step employed by Uneyama and Doi. The transformation affords a systematic parameterization of the free energy that preserves key structural features such as scattering structure factor.  A simple choice of transformation is shown to incorporate the Tuebner and Strey microemulsion structure factor and provide a reduction to the microemulsion free energy. Without adjustable parameters, the associated  phase diagrams are compared to experimental and self consistent mean field based results. A gradient descent of the free energy recovers dependence of end-state morphology on initial configurations, and identifies coexisting microstructures and transitions to two-phase behavior. Small angle x-ray data is simulated and used in classification of microphase morphology.  
\end{abstract}

\section{Introduction}

Blends of copolymers, polymers and solvents form a rich energy landscape with a complex phase diagram that affords remarkable potential for nano-scale control of self-assembling morphology \cite{Tree-23}.  Polymer filtration membranes are a particular target with important applications to water purification, biomedical separation processes such as protein purification and virus filtration, and pollution remediation \cite{MTrans1, MTrans2}. 
Conventionally, filtration membranes are designed through non-solvent induced phase separation \cite{ Memtrans3, Memtrans4}.  Recently self-assembled solutions of flexible polymer surfactants,  sometimes called lyotropic liquid crystals (LLCs) \cite{Foud-21}, have been used as templates for isoporous membranes which target enhanced filtration selectivity \cite{Foud-19, LLC2, LLC3, LLC4, LLC5, LLC6}. Blends of water, monomer, and block copolymers form a template from which the monomer phase is polymerized, crosslinking the film to produce a polymerized filtration membrane. The LLC-based process has demonstrated significantly higher selectivity than conventional membranes of similar pore size \cite{Foud-19, Foud-22}, with the ability to separate solutes with slight size differences and excellent anti-fouling properties due to the hydrophilicity of the pore lining \cite{Foud-21b}. 

Neat blends of AB diblock copolymers with weak AB segregation strength reside in a disordered state.  Increasing the segregation strength pushes the system through the order disorder transition (ODT) to form lamellar, hexagonal, gyroidal, and micellular microphases. More complex blends can develop microemulsion phases and other bicontinuous morphologies that do not have obvious symmetries. Bicontinuous phases have elicited substantial interest as templates for bound electrolyte phases in batteries, \cite{muE-Tutorial} and selective ionic conductors \cite{Hayward14}. However microemulsions can be particularly sensitive to auxiliary processing steps, being susceptible to shear-induced macrophase separation \cite{Bates-PRL01}.

 The map from initial composition of polymer blend to an end state, cross-linked polymer membrane is complex.   The final structure depends sensitively on a variety of material and process parameters. The inclusion of multiple homopolymers and copolymers greatly increases the complexity of fabrication, as microphase and macrophase separation must be properly organized on both length and time scales to obtain a membrane with the desired structure, \cite{Muller-CR21}.  Microphase domains can co-exist, \cite{Bates-Lodge16}, polydispersity in diblock copolymers and homopolymers can lower the order-disorder transition (ODT) point, stabilize microemulsions \cite{Manha10, Bates09, Hillmyer07}, and introduce longer-range interactions that interrupt ordered patterns and reduce translational symmetry, \cite{Mahan17}. The inclusion of charged groups \cite{Muth-22} and dissolved salts increases the experimental toolbox but further complicates the energy landscape. Indeed varying the salt concentration significantly impacts the effective segregation strength of polymers and can induce a full range of microphase bifurcations, \cite{Salt2}. 

Self consistent mean field models (SCMF), field theoretical models, and dissipative particle dynamics (DPD) approaches have amply demonstrated their ability to efficiently capture the equilibrium nature of polymer self-assemblies on tens of nanometers \cite{Fred-06, Dorfman-2016, DPD1, DPD2, Matsen20}. However simpler phase-field models can be helpful to scan large state spaces or facilitate the resolution of larger (hundreds of nanometers) length-scales associated to grain domains with distinct microphase separation. Moreover complex blends often require auxiliary processes such as thermal and shear annealing to eliminate defects, enlarge grain size, and coax the system into ordered states \cite{Epp14, Russell-AM14, Russell-Macro16}. Auxiliary processes pose two challenges to computational approaches. First they are non-equilibrium and require coupling to external fields. Their simulation requires a model for forced-base dynamics and introduces issues of system frustration and dependence on initial data.  Second, time scales of auxiliary processes and can be long compared to other system processes.  Some of these issues have been addressed for SCMF computations by incorporating variable cell shapes \cite{Fredrickson05} and screening from random initial data \cite{Fred-99}. However SCMF does not trivially allow coupling to auxiliary fields and does not always scale well to larger domains \cite{Dorfman-2016}. Recent work shows that gradient flows of phase field models have the potential to couple phase separation to kinetic fluid flow effects, \cite{Rheology-23, Fred-SM22}.

Density functional (DF) or phase-field models derived under the nearly spatially-uniform density assumptions of the random phase approximation (RPA) date back to Leibler \cite{LL-80} and Noolandi \cite{Noolandi-83}. The widely recognized Ohta-Kawasaki (OK) model \cite{OK-86} for neat diblock copolymers successfully predicts the ODT \cite{Glaser14} and salient features of the phase diagram. Generalizations of the OK free energy extended it to multicomponent blends \cite{CR-03, uneyama-05}. However the RPA introduces several uncontrolled approximations, for more complex blends can only hope for qualitative predictions for the resulting DF models. The Flory-Huggins parameters shape key aspects of these energies,  but they may not be sufficient to control all behaviors asked of them \cite{BLF-98}. Measurements of Flory-Huggins parameters are known to be broadly sensitivity to system properties, reproducibility requires carefully selected experiments \cite{Matsen20}.

Given this context, it may be  more realistic to treat RPA-based density functional models less as ab initio predictive tools and more as templates to be trained to data extracted from local conditions \cite{An-Chang20, Carlos-19}. It is particularly promising to drive the training from machine-learning based force-matching from particle or field based models, such as coarse-grained molecular dynamics and kinetic theory \cite{Lei_E_DeePN2_2022, Lei_Wu_E_2020, Ge_Lei_JCP_2023, Zhang_DeePCG_JCP_2018}. Machine-learning based training for condensed matter systems can be remarkably successful when the training possesses is preconditioned with a relatively low-dimensional template that embed key features of the system \cite{Liu-17}.  


This work presents a modification of the RPA derivation of density functional models that systematically builds in parameterization while preserving key structural components. As derived by Ohta and Kawasaki, the density functional  models incorporate long-range interaction terms describing the entropic effects of chain folding and volume exclusion derived from the Feynman-Kac relations, but only to a linear level. The ``strong-segregation extrapolation'' (SSE) step incorporates essential higher-order terms into the model but does so in an ad-hoc manner. Specifically it incorporates singular structure into the entropic terms in a manner that preserves positivity of the density functionals in the strong segregation limit.  Generically the form of the SSE step has been chosen so as to preserve the scattering structure near the order-disorder transition. The parameterization step incorporates nonlocal transformation of the density functions associated to the amphiphilic polymers prior to the SSE step. This provides a wide range of models which preserve the structure factor. For clarity, the parameterized density functional (pDF) model is presented for a particularly simple nonlocal transformation with a single free parameter. The resulting pDF increases bistability and makes a connection to the scalar microemulsion models of Gompper and Kraus \cite{Gompper-K_I-93} in a wet-brush regime.  In this sense the parameterization approach provides a more systematic connection between the RPA approximations of homopolymer/homopolymer/diblock copolymer blends and the microemulsion models developed for water/oil/surfactant systems, in agreement with the analogy of Bates and Lodge, \cite{Bates-Lodge16}.

\section{Parameterized Density Functional Model}

The mean field and RPA traditionally address a blend of diblock copolymers, homopolymers, and solvent described by their $M$ volume fractions  $\{\phi_i\}_{i=1}^M$ on a physical domain $\Omega=[0,L]^3$ subject to periodic boundary conditions. The index $i$ ranges over each distinct polymer or solvent species, including each block component of the diblock chains.
The free energy is expressed in terms of external fields $U=\{U_i\}_{i=1}^M$ and the volume fractions.
Within the mean field theory the intra-molecular interactions are described by the product of the volume fractions and the external fields, while the inter-molecular interactions are governed by the usual Flory-Huggins relations,
 \begin{equation}\label{eq:Fdef}
\mathcal F(\phi)=\int_\Omega \left(\sum_{i,j} \chi_{ij}\phi_i\phi_{j}-  \rho_0\sum_i U_i(x) \phiro_{i}\right)\, dx. 
\end{equation}
 The volume fractions have been decomposed into their domain averages $\{m_i\}_{i=1}^M$ and the deviation $\{\phiro_i\}_{i=1}^M$, which satisfy 
$\phi_i=m_i+\phiro_i$. 
The system is subject to the incompressibility condition
$
\sum_{i} \phi_{i}=1.$

As critical points of the SCMF energy the external fields and volume fractions are related through the forward and backward propagators which describe the conformations of the polymer chains under the influence of the other system components. Following the approach of Uneyama and Doi the first step of the RPA approximation, outlined in the supplementary material,  derives a linear approximation to the propagator relation. This employs the long-chain approximation for the Fourier symbol for the propagator relations. These relations are diagonal over each {\sl connected} polymer component. Substituting the linear relation of the external field on the volume fractions into the free energy yields a quadratic characterization of the weak segregation regime, 
\beq
\label{eq:UD2}
\cF^{(2)}_{\rm }(\phiro) = \sum_{i,j}\int_\Omega 
 \frac{a_{ij}}{m_i} |\nabla \phiro_{i}|^2 +
\left(\frac{b_{ij}}{\sqrt{m_im_j}}+\chi_{ij}\right)\phi_{i}^0\phi_{j}^0 +
\frac{c_{ij} (\mrG\phiro_{i}) (\mrG\phiro_{j})}{\sqrt{m_i m_{j}}} \,  \textrm{d}x.
\eeq
The operator $\mrG:=(-\Delta)^{-\frac12}$ is the square-root of the negative inverse Laplacian subject to periodic boundary conditions. Length has been scaled by the Kuhn length $\ell$ and energy by $\ell^3$. The interaction matrices $a=(a_{ij})$, $b=(b_{ij})$, $c=(c_{ij})$  are block diagonal over connected polymer components and have entries that depend only upon the chain structure: specifically the chain lengths of homopolymers and then lengths of each copolymer block in a diblock chain.  For homopolymers the $c$ term is zero, and for solvent phases both $a$ and $c$ are zero. Explicit forms are given in the supplementary material. 





\subsection{Uneyama-Doi model from the Strong Segregation Extrapolation}
The  SSE is the least controlled step in the energy development is the extension to a nonlinear model. The approached taken by Uneyama and Doi introduces singularities the preserve positivity of the densities. This is crucial in the strong segregation limit, $\chi N$ large, for which phase separation drives some volume fractions to zero. Significantly this extrapolation step is not unique.  Uneyama and Doi replace the bulk densities $m_i$ with the spatially varying densities $\phi_i$,
 $$ \bphi_i\mapsto \phi_i=\bphi_i+\phiro_{i}$$
 while imposing the condition that the extrapolated model $\cF_{*}$ preserves the structure factor found in the bilinear form of the energy. This is equivalent to imposing the condition
\beq\label{e:StucFact} \frac{\delta^2 \cF_{*}}{\delta \phi^2}\Bigl|_{\phi=m}= \frac{\delta^2 \cF^{(2)}}{\delta \phi^2}\Bigl|_{\phi=m}.
 \eeq
This forces the extrapolated energy to agree with the bilinear form at the ODT. There are a host of extrapolated models that satisfy this relation. Uneyama and Doi make the choice \cite[eqn.(46)]{uneyama-05} which  in the current notation  takes the form
\beq
\begin{aligned} 
\cF_{\rm UD}(\phi) &= \int_\Omega \sum_{i}\left( \frac{ a_{ii}}{\phi_i} |\nabla \phi_{i}|^2 + b_{ii} \phi_i\ln\phi_i\right) + \sum_{j\neq i}
\left(\frac{b_{ij}}{\sqrt{\phi_i\phi_{j}}}+\chi_{ij}\right)\phi_{i}\phi_{j}\\
&\hspace{0.4in} + \sum_{i,j}
\frac{c_{ij}}{\sqrt{\phi_i\phi_{j}}} \mrG (\phi_{i}-m_i)\mrG (\phi_{j}-m_j)\,  \textrm{d}x.
\end{aligned}
\eeq
This energy functional give quantitative agreement with the scattering functions for AB diblock polymers \cite[Figure 2]{uneyama-05}. It reduces to the Flory-Huggins-de Gennes type free energy for homopolymer blends and to the Ohta-Kawasaki energy for a  neat diblock copolymer melt. It inherits the sharp spinodal decomposition mechanism, \cite{LL-80}, of the models of Flory and Huggins, \cite{Flory-42, Huggins-42} that preclude the existence of microphase separation below the ODT line and recovers the strong quantitative agreement of the phase diagram for neat AB diblock copolymer melts.

\subsection{Parametrized SSE}
We consider an alternate SSE step which emphasizes the scattering structure for microemulsions introduced by Tuebner and Strey \cite{TS-87}.   To simplify notation the presentation is focused on a binary blend of an AB copolymer diblock and an A homopolymer.
Disorder near the diblock-homoploymer interface is systematically introduced via a nonlocal change of variables (spatial averaging) imposed on the amphiphilic diblock volume fraction prior to the SSE. The change of variables is motivated by the propensity for amphiphilic interfaces to spontaneously disorder, for example in the lamellar-to-disorder transition  which may yield a microemulsion phase, \cite{Bates14}. Diblock copolymers residing at the interface phases are susceptible to longer range correlations over hundreds of nanometers. These effects can arise from copolymer polydispersity, which has the tendency to disrupt order, partial charge effects between amphiphilic moieties and solvent that have a propensity to induce longer range order. 
 
For the binary blend with densities $(\phi_A, \phi_B, \phi_H)$ the parameterization is induced through a change of variables transformation imposed only on the AB diblock copolymer volume fractions $(\phi_A,\phi_B)$. These are spatially averaged through an operator $\cG$, yielding the locally averaged volume fraction
\beq
\label{e:eDF_G}
\psi_i:=  \bphi_i + \cG\phiro_{i}=
 m_i + \cG \left(\phi_{i} -m_i \right), 
\eeq 
for $i=$A,B. In general $\cG$ can be any positive averaging operator that is an invertible  map from zero-mass densities to zero-mass densities. This includes the  Fourier operators (convolutions) with symbol $\widehat{\cG}$  and multiplicative Fourier action given by
$$ \widehat{\cG \phiro}(\zeta)= \widehat{\cG}(\zeta)\widehat\phiro(\zeta),$$
so long as $\widehat\cG(\zeta)\geq0$ and any singularity at $\zeta=0$, satisfies $\widehat{\cG}(\zeta)|\zeta|$ bounded as $|\zeta|\to0.$ 
 The identity \eqref{TS-id} allows the quadratic energy for a diblock copolymer \eqref{eq:UD2} to be written in the \emph{equivalent} local formulation,
\beq
\label{eq:cGS2}
\cF^{(2)}_{\rm AB }(\psi) =\hspace{-0.15in} \sum_{i,j\in\{A,B\}}\int_\Omega 
 \frac{a_{ij}}{m_i} |\Delta \cG^{-1}\psi_{i}|^2 +
\left(\frac{b_{ij}}{\sqrt{m_im_j}}+\chi_{ij}\right)\cG^{-1}\psi_{i} \cG^{-1}\psi_{j} +
\frac{c_{ij} (\cG^{-1}\mrG\psiro_i) (\cG^{-1}\mrG\psiro_{j})}{\sqrt{m_i m_{j}}} \,  \textrm{d}x.
\eeq
Here $\cG^{-1}$ is understood to map onto zero-mass densities, and $\psiro_i=\psi_i-m_i$ has zero mass. To this point the parametric approach is merely a change of variables that yields a quadratic energy equivalent to the original. The divergence from the UD approach occurs at the SSE step. This involves the replacement of the bulk volume fractions of the diblock-copolymer/surfactant phases, $m_i$, $i=A,B$ by the averaged volume fraction $\psi_i$ rather than by the local volume fraction $\phi_i$,
 \beq
    \bphi_i \mapsto \psi_i\hspace{1.0in} i={\rm A, B}.
  \eeq 
Making this replacement in \eqref{eq:cGS2} is not equivalent to making a replacement in \eqref{eq:UD2} followed by the change of variable. Moreover the replacement is made only in the second ($b$) and zero-th order ($c$) terms as this increases their ability to induce bistability.  The result is the parametric formulation for the diblock copolymer self-interaction,
 \beq\label{eq:cTS}
\cF_{\rm pAB}(\psi)=\hspace{-0.15in}\sum_{i,j\in\{A,B\}}\int_\Omega
 \frac{a_{ij}}{m_i} |\nabla \cG^{-1} \psi_{i}|^2 +
\left(\frac{b_{ij}}{\sqrt{\psi_i\psi_{j}}}+\chi_{ij}\right)\cG^{-1} \psi_{i}\cG^{-1}\psi_{j} +
\frac{c_{ij}(\cG^{-1}\mrG\psiro_i)(\cG^{-1}\mrG\psiro_j)} {\sqrt{\psi_i\psi_{j}}} \,  \textrm{d}x.
\eeq
The parameterized density function ($\eDF$) energy is formed by coupling the energy \eqref{eq:UD2} for the homopolymer $\phi=\{\phi_H\}$ with parametric diblock model for
$\psi=\{\psi_A,\psi_B\}$ through the Flory-Huggins interactions
\beq\label{e:pDF}
\cF_{\rm pDF}(\psi,\phi)= \cF_{\rm pAB}(\psi)+\varrho_A\int_\Omega 2\chi_{AB}\phi_H\cG^{-1}\psi_B\, \,\textrm{d}x + \cF_{\rm UD}(\phi).
\eeq
The structure factor condition \eqref{e:StucFact} holds for $\cF_{\rm \eDF}$ for any of the choices of $\cG$ outlined above since $\cG^{-1}m_i=0$ and the second variational deriviative must act on the $\cG^{-1}\psi_i$ terms to have non-zero impact. 
The $\eDF$ model reduces to the Uneyama-Doi model for the choice $\cG=\mathrm{I}$.
The incompressibility constraint is imposed on the mixed variables
\beq\label{e:incomp}
\psi_A+\psi_B+\phi_H=1,
\eeq
as those are the quantifies that the energy keeps positive.

The sequel of this work addresses the specific choice
$ \cG   =\frac{1}{\La}\mrG$
 of averaging through a scaling of the nonlocal operator $\mrG$ that arises in the long-chain approximation.
 The parameter $\La$ modifies the  length scale of the averaging operator and makes $\cG$ non-dimensional.
 For this choice the averaging operator has Fourier symbol
$$ \widehat{\cG}(\zeta) = \frac{1}{\La|\zeta|}.$$ 
 This yields the simplest formulation that has a direct connection to the Tuebner-Strey energies for microemulsions.  Moreover, while the inverse operator $\mrG^{-1}$ is nonlocal its quadratic form has a local formulation
\beq \int_\Omega (\mrG^{-1} \psi_i)
\label{TS-id}
(\mrG^{-1}\psi_j)\,\textrm{d} x =\int_\Omega (-\Delta\psi_i)\psi_j\,\textrm{d}x= \int_\Omega \nabla\psi_i\cdot\nabla\psi_j \,\textrm{d}x.\eeq

Scaling space by $\ell$, energy by $\ell^3$, and employing the identity \eqref{TS-id} the  diblock copolymer energy \eqref{eq:cTS} takes the scaled, local $\mrG$-formulation,
 \beq
 \label{e:TS}
\cF_{\rm GAB}(\psi) =\epsA^2\hspace{-0.15in}\sum_{i,j\in\{A,B\}}\int_\Omega
 \frac{a_{ij}}{m_i} |\nabla^2 \psi_{i}|^2 +
\left(\frac{b_{ij}}{\sqrt{\psi_i\psi_{j}}}+\chi_{ij}\right)\nabla \psi_{j}\cdot\nabla\psi_{j'} +
\frac{c_{ij}(\psi_i-m_i) (\psi_{j}-m_{j})} {\sqrt{\psi_i\psi_{j}}} \,  \textrm{d}x.
\eeq
The dimensionless parameter $\epsA=\La/\ell$ is the ratio of averaging length to the Kuhn length.
For this averaging the coupling to the homopolymer 
$\phi=\{\phi_H\}$ yields the $\mrG$ formulation of the DF model
\beq\label{e:GDF}
\cF_{\rm GDF}(\psi,\phi)= \cF_{\rm GAB}(\psi)+\varrho_A\int_\Omega 2\chi_{AB}\phi_H\mrG^{-1}\psi_B\, \,\textrm{d}x + \cF_{\rm UD}(\phi).
\eeq
An explicit formulation, \eqref{eq:AB-H}, is given in the supplementary material. The structure factor condition \eqref{e:StucFact} holds for $\cF_{\rm \GDF}$ for any of the choices of $\cG$ outlined above as the averaging operator has no impact on spatially uniform densities.  

For blends with multiple types of diblock copolymers each co-polymer and homopolymer could have a distinct averaging parameterization. The $\mrG$-choice amounts to no averaging for the $A$-homopolymer and averaging by scaled $\mrG$ for the diblock. This results in Flory-Huggins type energy for the homopolymer  which encourages macrophase separation, while the gradient terms in the $\cF_{\rm GAB}$ energy reward the generate of interface.  The dimensionless parameter $\epsA$ scales the competition between these two effects. The interaction between the diblock $\psi$ and homopolymer $\phi$ components occurs solely through the $\varrho_A$ scaled Flory-Huggins term $\chi=\chi_{AB}$.  

\section{Computational Results}

For binary blends of AB diblock and A homopolymer, the incompressibility constraint allows the energy to be written as a function of $\psi=(\psi_A,\psi_B)$.  The domain $\Omega=[0,L]^3$ is periodic with length $L=100$nm set by a one-time scaling to set the $f_A=0.5, m_H=0$ lamellar width at 10nm, see Figure\,\ref{f:SAX}(b). To facilitate comparison to experimental data in \cite{TwoSolvBD_Bates} the diblock chain length is fixed at $N_{AB}=N_{A}+N_{B}=86$ and homopolymer of chain length $N_{H}=16.$

For models with bistabilty it is important to understand the basin of attraction of competing stable states. This requires construction of the map from initial configuration to final equilibrium. Within the Onsager formulation for an overdamped system, \cite{Ons31, Ons31b} this requires to specify the dissipation mechanism of the system. The dissipation operators must be positive and annihilate the constants so that the energy decays and the system mass is preserved.  The choice $\cD=-\frac {\Delta}{1-\Delta}$ has positive Fourier symbol $\hat\cD(\zeta)=|\zeta|^2/(1+|\zeta|^2)\geq 0$ with $\hat\cD(0)=0$, and hence annihilates constants.  It induces a  uniform dissipation rate for length intermediate to short length scales, (large $|\zeta|$). The associated gradient flow takes the form
 \beq
 \label{e:GF}
 \psi_t = -\cD\frac{\cF_{\rm \GDF}}{\delta \psi} =  \frac {\Delta}{1-\Delta} \frac{\delta \cF_{\rm \GDF}}{\delta \psi},
 \eeq
 subject to periodic boundary conditions.

 Phase diagrams are a key benchmark for computational models. A traditional and efficient route to their computation is to obtain characterizations of candidate phases for energy minimizers and extend them through numerical continuation over their basin of stability. Where multiple candidate minimizers coexist, their energy levels are compared, with the lowest energy selected as the dominant morphology for the given system parameters. While computationally efficient this equilibrium based approach has a few drawbacks. One is that it does not predict candidate phases, but requires them as input, \cite{Matsen20}. 
 \begin{wrapfigure}{l}{0.4\textwidth}
   \centering
\includegraphics[width=0.3\textwidth]{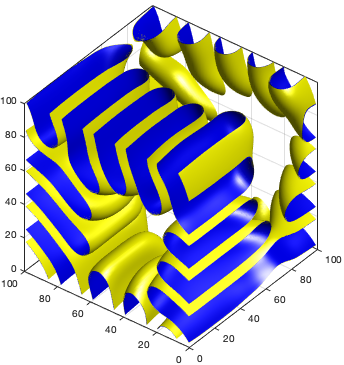}
\vspace{-0.0 in}
    \caption{\small Two level sets of the seeded initial data of $\psi_B$ at 50\% (yellow) and 0.52\% (blue) of its maximum value.}
    \label{f:IC}
    \vspace{-0.15in}
\end{wrapfigure}A second is that as system complexity increases achieving a desired equilibrium from experimentally accessible initial configurations may require auxiliary processing, and this is easier to incorporate into a non-equilibrium gradient flow simulation.  Indeed the incorporation of random initial data to improve the sampling process has been considered, \cite{Fred-99}. 

 Another option is seeded initial data based upon a non-equilibrium initial configuration which combines certain structural elements. In the majority of the simulation results presented the initial data is seeded, based upon the fixed configuration presented in Figure\,\ref{f:IC}. The initial data is modified only to have desired mass fractions by adding spatial constants to the initial data and then truncating volume fractions to lie within  $[0,1]$. This particular choice of seed, with both lamellar and void regions, stimulates competition between microphase domains on the computational lengths available.  In some 
simulations initial data is derived from spatially uniform volume fractions perturbed by $10\%$ zero-mass white noise, or are formed from computational continuation of equilibrium as in the study of hexagonally packed cylinders. 

A choice of dissipation operator implies a scaling of time. The onset of the spinodal decomposition of a polymer blend is a fundamental benchmark for time scales.  Dynamic light study of the onset of the spinodal decomposition in a long chain $N\approx 10^3$ polystyrene/polyvinyl methyl ether blend shows time scales on the order of 30 minutes for the development of the initial spinodal decomposition \cite{Hash-83}.  For the gradient descent simulations, the ratio of final simulation time to onset of spinodal decomposition from perturbed uniform initial data is generally 1000:1.  The homopolymer chains in the experimental study are substantially longer than those simulated which suggests a slower spinodal decomposition, however the rough estimate of computational time is on the order of 200-500 hours of laboratory time. Doubling the time of simulations generically yielded visually undetectable changes to structure and small changes to system energy.

A key feature in binary (AB+A) and ternary (AB+A+B) diblock copolymer/homopolymner blends is the transition from a wet-brush regime, in which a homopolymer swells the like-monomer block of the diblock, and a two-phase regime in which they separate to form homopolymer-rich domains. This transition was investigated in foundational work by Hasimoto's group. Conducting small-angle X-ray (SAX) scattering on diblock-homopolymer binary and ternary blends they found that low-weight homopolymers $(\alpha<0.25)$ were completely solubilized by the associated diblock (wet brush) \cite{Hash-90-2}. Increasing homopolymer chain length induced two-phase behavior and longer range order in the SAX measurements indicated by small $q$ upticks in the scattering intensity \cite{Hash-90-1}. For simulations it is convenient to characterize two-phase behavior through the $L^2(\Omega)$ projection of the homopolymer volume fraction onto the volume fraction of the $A$-phase of the diblock,
$$\mrI_{\rm AH} := \frac{\langle \psi_A,\phi_H\rangle}{\|\psi_A\|\|\phi_H\|}.$$
This quantity takes values in $[0,1]$ and tracks the percentage of homopolymer participating in swelling the $A$ block of the diblock copolymer. In particular $\mrI_{\rm AH}$ approaches $1$ in the wet brush regime when the $A$-diblock and $A$-homopolymer volume fractions  interdigitate, with their densities becoming multiples of each other, $\phi_H= (m_{H}/m_{A})\psi_A.$ A two-phase blend is defined by the condition $\mrI_{AH}<0.95$, as this threshold generically coincides with the development of spatial domains of pure homopolymer.

\begin{figure}[h!]
\includegraphics[width=0.55\textwidth]{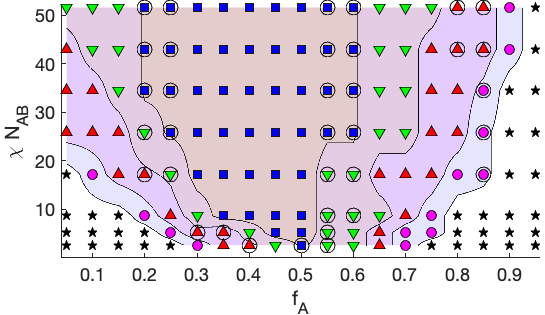}
\vspace{-0.1 in}
    \caption{\small Dominant microphase morphology in the AB-A binary blend from seeded initial data as function of diblock fraction $f_A$ and segregation strength $\chi N_{AB}$ for $\alpha_A=.186$, $m_H=0.2$ and $\epsA=1.$  For fixed values of $\chi N_{AB}$ the microphase morphology transitions from disordered (black star) to micelles (magenta circles), to hexagonally packed cylinders (red triangles) to gyroidal phase (green inverted triangles) to lamella (blue squares), and then back with increasing values of $f_A.$ A circle around the symbol indicates a wet brush with coexisting microphase domains as determined algorithmically by condition \eqref{e:Small-q}. The dominant morphology within the two-phase structure is indicated. Lines separating regions of common dominant morphology are included to guide the eye.}
    \label{f:BD-alpha}
\end{figure}


\subsection{Phase Diagrams}

 SCMF models are well known to give accurate predictions of the $f_A-\chi N$ phase diagram for conformationally symmetric diblock melts, \cite{Matsen96}. While fine details of this phase diagram have been debated, \cite{Carlos-06}, it presents the order-disorder transition and the selection of microphase morphology of minimal energy as a function of diblock fraction $f_A=N_{A}/N_{AB}$ and of the segregation strength $\chi N_{AB}.$ The phase diagram is symmetric about the $f_A=0.5$ line, and above the spinodal line, one finds lamellar, double-gyroid, hexagonally packed cylinders, close packed spherical phases, and disordered mixtures emerging with increasing deviation of $f_A$ from 1/2.  The DF model of Uneyama and Doi provides a quantitatively accurate reproduction of the SCMF phase diagram \cite{uneyama-05}[Figure 4].

\begin{figure}[h!]
   \centering
    \includegraphics[height=2.5in]{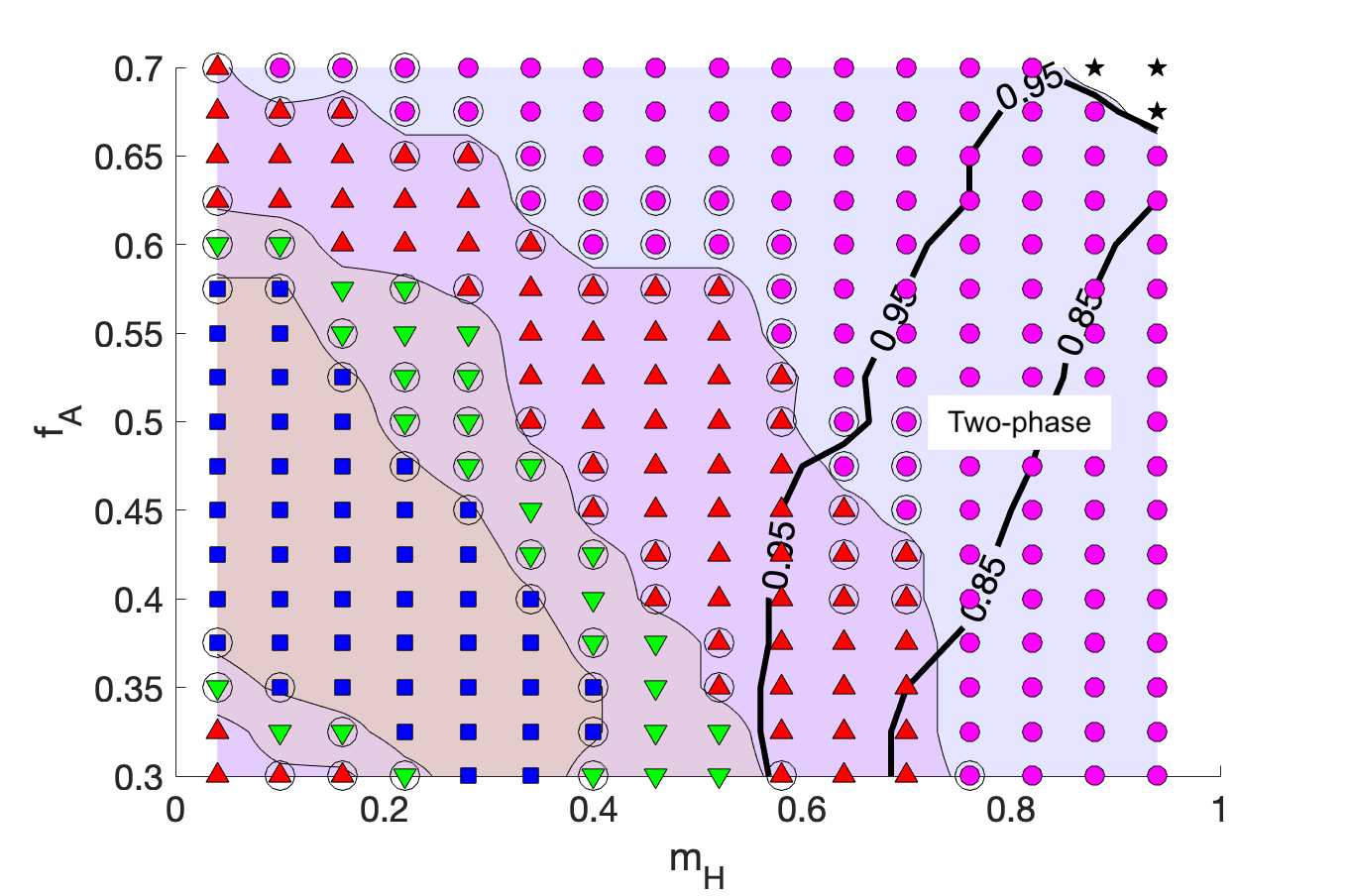}   \includegraphics[height=2.4in,width=2.6in]{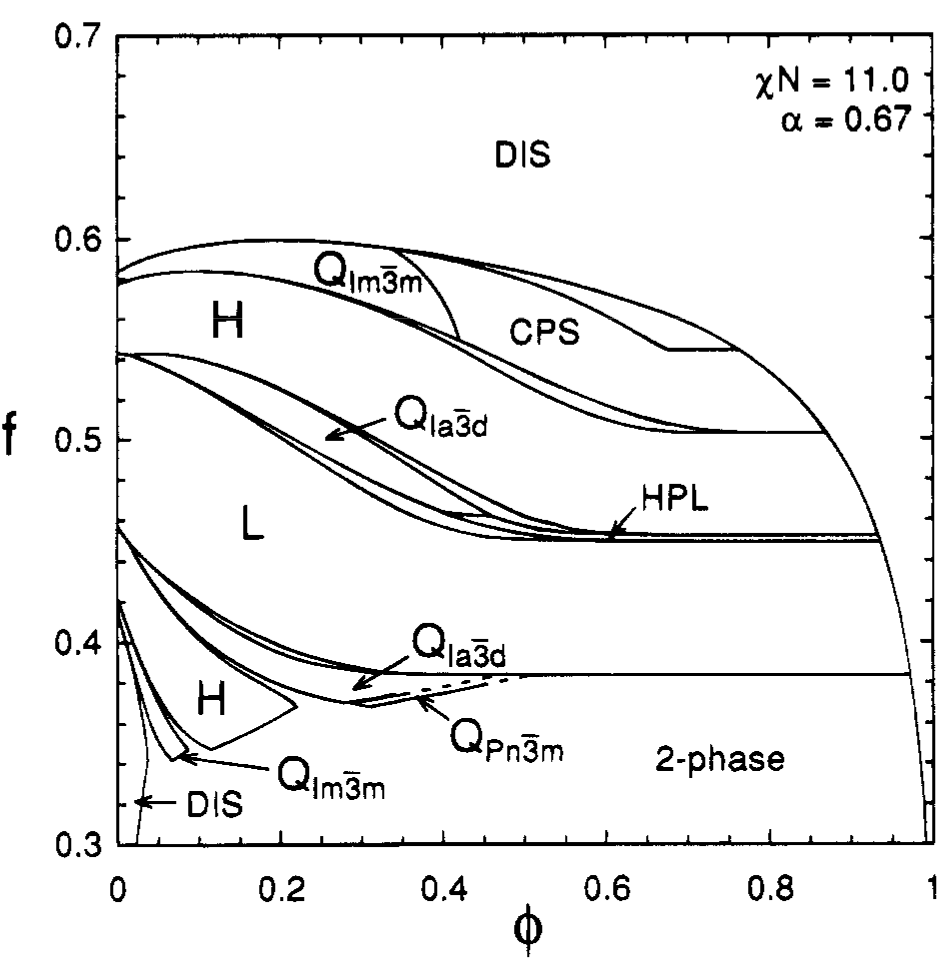}
   \put(-465, 160){\Large\textbf{(a)}} 
   \put(-210, 160){\Large\textbf{(b)}}
   
    \caption{Comparison of the $m_H-f_A$ phase diagram for a binary AB-A blend. The homopolymer to diblock chain length ratio $\alpha=0.67$ and the segregation parameter $\chi N_{AB}=11$ are fixed. (a)  End states of a gradient descent from seeded initial data of the microemulsion energy with $\rho_A=0.67$ (symbols as in figure\,\ref{f:BD-alpha}). Contours of $\mrI_{\rm AH}$ at 0.85 and 0.95 (thick black), two-phase region is marked.   (b)  Results of computational continuation from the self-consistent mean field model for similar binary polymer blend with L=lamellar, H=hexagonal, Q$_{\rm la\overline{3}d}$=gyroidal, and Q$_{lm\overline{3},m}$=micellular phases. (reprinted from \cite{Matsen95}[figure 8]).}
        \label{f:BD-fA}
\end{figure}

As a benchmark comparison, the $f_A-\chi N_{AB}$ phase diagram for a binary blend of diblock copolymer with $20\%$ volume fraction of homopolymer of phase A ($m_H=0.2$) was computed from the gradient flow \eqref{e:GF} of the parameterized DF model with seeded initial data.  
 The diagram is asymmetric, shifted to the left about the $f_A=0.5$ mid-line reflecting the symmetry breaking influence of the 20\% volume fraction of the homopolymer, see Figure\,\ref{f:BD-alpha}. The ratio of homopolymer to diblock chain length $\alpha_A=N_{H}/N_{AB}=.186$ is below the threshold at which two-phase behavior is anticipated, \cite{Hash-90-1, Hash-90-2, Schick-97}. The $\GDF$ gradient flow end states are all in a wet-brush regime with the A homopolymers swelling the A block of the diblock copolymer. Indeed the diblock-homopolymer projection satisfies $\mrI_{\rm AH}>0.97$ over all simulations. The key deviations of the $\GDF$ phase diagram lie in the blurring of the ODT transition, in particular microphase separation persists below $\chi N_{\rm AB}<3$.  For spatially uniform phases, $\psi_i=m_i$, the energy $\cF_{\rm pAB}$ model is zero, independent of the Flory-Huggins parameters.  In the $\GDF$ model the spinodal line merely indicates the onset of instability of the disordered phase, but the $\cF_{\rm \GDF}$ model does not preclude the existence of stable microphase separation below the line. While the ODT is a hallmark of binary homopolymer blends, the overlapping of stability domains is not uncommon in multi-component blends,  \cite{Schick-97,TwoSolvBD_Bates}. The gradient descent from seeded initial data actively encourages coexistence of competing microphase domains in the end state structure, as indicated the phase diagrams by a circle about the phase symbol. In Figure\,\ref{f:BD-fA} the coexisting domains are included mixed lamellar and cylindrical structures as have been reported in experiments with Li-salted polystyrene-poly(ethylene oxide) (PS-PEO) ternary blends, \cite{Bates-Lodge16}[Figure 4].

\

 The $f_A-m_H$ phase diagram for a binary AB-A blend fixes the homopolymer to diblock chain ratio, $\alpha_A$, and segregation strength $\chi N_{AB}$ and varies the homopolymer volume fraction $m_H$ and diblock chain fraction, $f_A$. 
 The phase diagram has two central features: the regions of distinct microphase morphology and the onset of two-phase behavior, in which homopolymer segregates from the diblock to form homopolymer rich domains. Taking $\alpha_A=0.67$ and $\chi N_{AB}=11$ the results of gradient flows of the $\GDF$ model from seeded initial data and $\rho_A=0.67$ are presented in Figure\,\ref{f:BD-fA} (a) and compared to computational continuation results from the self-consistent mean field model reprinted from \cite{Matsen95} in Figure\,\ref{f:BD-fA} (b).  
 The SCMF computational continuation  reveals a complex phase structure with the micelle-hexagonal-gyroidal-lamellar transitions sandwiched between a two-phase regime for $f_A$ (labeled $f$ in Figure\,\ref{f:BD-fA}(b)) below 0.4 and a disordered regime for $f_A$ above 0.6. The onset of two-phase behavior is strongly influenced by the value of $f_A.$
 The $\GDF$ phase diagram yields qualitatively comparable results for smaller values of $m_H$ (labeled $\phi$ in subfigure (b)), but with larger regions of lamellar and gyroidal structures as well as coexistence of microphase domains (indicated by circle around symbol) near the phase transition boundaries. 
 In Figure\,\ref{f:BD-fA}(a) the boundary of the two-phase domain is indicated by the contour $\mrI_{\rm AH}=0.95$. The  contour $\mrI_{\rm AH}=0.85$ is included to indicate direction and magnitude of the gradient.  The two-phase region in the $\GDF$ phase diagram arises at larger values of $m_H$ and smaller values of $f_A$. It occupies  the lower-right hand corner except for a truncation in the upper right where it abuts the disordered phase, for which $\mrI_{\rm AH}=1.$ The disordered phase is much smaller in the $\GDF$ simulation, and is largely replaced with a micellular regime. The seeded initial data strongly enhances the probability that a gradient flow computation terminates on a microphase structure even when the disordered state is stable.  
 
 For the $\GDF$ model, the sensitivity of the two-phase region to $m_H$ implies that the two-phase behavior is triggered by the displacement of the homopolymer chains from the microphase structure in an unbinding transition. The energy cost of  homopolymer beyond the unbinding limit is dominant and the excess is ejected, forming homopolymer rich domains. It has been suggested that highly swollen structures may have marginal stability \cite{Matsen95}. This may lead highly swollen states to be favored by computational continuation over gradient flow.  
 The computational continuation approach for SCMF minimizers associates the two-phase regime to smaller values of $f_A,$ for which the $A$-block chains are shorter. This suggests the onset of the two phase regime may be driven more by entropic competition between homopolymer and diblock chains rather than by energetic limits of microphase swelling leading to an unbinding transition.
 

\begin{figure}[h!]
   \centering
    \includegraphics[width=2.95in, height=1.5in]
    {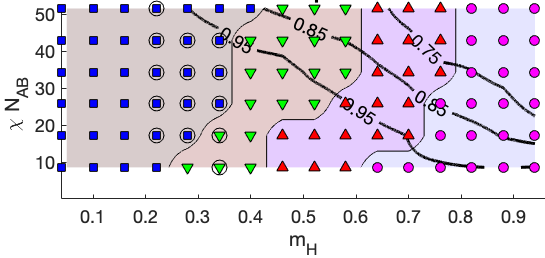}
    \hspace{0.1in}
 \includegraphics[width=2.95in, height=1.48in]
    {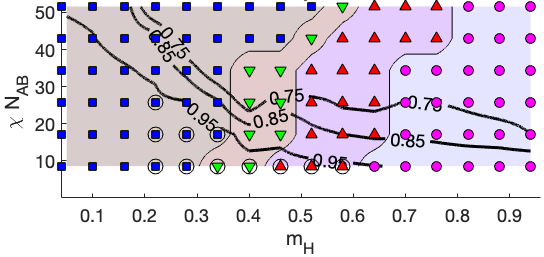}
          \put(-452, 95){\Large\textbf{(a)}}   \put(-226, 95){\Large\textbf{(b)}} 
    \caption{Dominant morphology in the AB-A binary blend and contour lines of $\mrI_{\rm AH}$ as function of homopolymer fraction $m_H$ and $\chi N_{AB}$ for seeded initial data with $f_A=0.5$ and (a) $\epsA=1$ and (b) $\epsA=1.67$. The level sets of $\mrI_{\rm AH}$ are superimposed in dark black, with two phase regions corresponding to $\mrI_{\rm AH}<0.95$. The dominant phase is labeled using symbols as in Figure\,\ref{f:BD-alpha}. }
    \label{f:BD-LamA1}
\end{figure}
The dimensionless parameter $\rho_A$ is the sole adjustable parameter considered in the $\GDF$ model considered herein. This parameter scales the diblock copolymer self-energy against the energy of its interaction with the homopolymer. In a force-field fitting approach its value will likely dependent upon many system parameters, including the polymer chain lengths. The impact of $\rho_A$ on the $m_H-\chi N$ phase diagram is presented in Figure\,\ref{f:BD-LamA1} for the $\GDF$ gradient descent from seeded initial data. 
For a symmetric diblock,   $f_A=0.5$, with  $\alpha_A=16/86\approx 0.186$,  Figure\,\ref{f:BD-LamA1} (a) presents the phase diagram for $\epsA=1.0$. The phases transition from lamellar to gyroidal to hexagonally packed cylinders and to micellular structures with increasing homopolymer content. These regions of each phase depend only modestly on values of $\chi N_{AB}> 20$. Each of the morphological transitions is gradual in $m_H$ with spatially coexisting morphologies observed for a range $m_h\approx \pm0.05$. As indicated in Table 1, these results have reasonable qualitative comparison to experimental observations. The study of PEO-PEP diblock with $f_A=0.5$ in a PEP homopolymer with $\alpha_A<0.25$, report a similar lamellar-gyroidal-hexagonal-micellular microstructure phase shifts at $100^\circ$C \cite{Bates-00}[Figure 6(b)]. 
The reported $m_H$ ranges for each phase of microstructure are largely independent of temperature over a range corresponding to $\chi N_{AB}\in[25,50]$. Lower temperature behavior can be extracted from the isothermal phase diagram at T=80$^\circ$C from \cite{Bates-00}[Figure 10]. The left edge of the triangle corresponds to a binary PEO-PEP plus PEO homopolymer blend. The change in temperature, and increase in effective $\chi$, induces shifts in the transition points and introduces a two-phase region at $m_H>0.90.$
Experiments of with a binary blend of a symmetric PEO-PS diblock copolymer with PS with a chain length $N_{PEO-PS}=86$ and $N_{H,PS}=18$ were conducted in a 6\% LiFTSI salt doped solution at $70^\circ$C.  The left edge of the isothermal phase diagram reported in \cite{TwoSolvBD_Bates}[8a]
shows similar qualitative phase transitions with a large region of disordered phase for $m_H>0.70.$
These experimental phase diagram results agree well with those from the $\GDF$ gradient flow with $\chi N_{AB}=8.6$, as reported in Table 1 (a, b, c, d). 

The impact of increasing $\rho_A$ to $\rho_A=1.67$ is presented in Figure\,\ref{f:BD-LamA1} (b). The transitions in microstructure phase are shifted incrementally, with the lamellar region invading the gyroidal region at strong segregation strength. This coincides with a left-ward shift of the two-phase region that signals the expulsion of homopolymer from the lamellar microstructure. For large values of $\epsA$ the lamellar microstructure has less energetic cost to eject excess homopolymer beyond is unbinding limit as the shift to $\epsA=1.67$ decreases the energetic cost of forming homopolymer-rich domains. The increase in $\epsA$ has significant impact on two-phase transition, roughly halving the value of $m_H$ at which $\mrI_{\rm AH}=0.95$ for fixed segregation strength. For both $\rho_A=1.0$ and $1.67$ the two-phase region grows with increasing segregation strength. Within the $\GDF$ model the unbinding point is lowers with increasing  strong segregation and microphase structure experiences increased cost to accommodate the intercalated homopolymer chains. 

\begin{center}
Table 1) Comparison of Phase Diagrams
\begin{tabular}{ |l|c|c|c|c|c|c| }
\hline
&Lamellar & Lam-Hex & Gyroid & Hexagonal & Micelle & Two-phase \\
\hline 
a) $\GDF$ with $\chi N_{AB}=8.6$ & $[0,0.25]$ & - & $[0.25, 0.45]$ & $[0.45, 0.65]$ & $[0.65,0.85]$ &$[0.85,1.00]$ \\
b) PEO-PEP+PEP/$100^\circ$C &   $[0,0.27]$ & - &  $[0.27, 0.37]$ & $[0.37, 0.60]$ & $[0.65,1.00]$ &  - \\
c) PEO-PEP+PEP/$80^\circ$C &   $[0,0.22]$ & - &  $[0.22, 0.40]$ & $[0.40, 0.67]$ & $[0.67,0.88]$ & $[0.88,1]$  \\
d) PS-PEO + PS & $[0,0.25]$ &-  &  $[0.25, 0.35]$& $[0.35, 0.75]$ & - & -  \\
\hline 
e) $\GDF$ with $\chi N_{AB}=34.4$ & $[0,0.20]$ &$[0.2,0.35]$ &$[.35, .50]$ & - & - &$[0.5,1.00]$\\
f) Ternary PS-PEO & $[0,0.50]$& $[0.50,0.70]$& - & - & - & $[0.75,1]$   \\
\hline 
\end{tabular}
\begin{tabular}{p{6.0in}}
\small{ 
 a) From $\GDF$ model with seeded initial data and $\chi N_{AB}=8.6$. Other parameters as in Figure\,\ref{f:BD-LamA1}(a).  
 b) From \cite{Bates-00}[Figure 6(b)], for PEO-PEP diblock with $f_A=0.5$ and PEP homopolymer with $\alpha_A<0.25$ at $100^\circ$C. Samples were thermally annealed. 
 c)  \cite{Bates-00}[Figure 10], same as line b) except at $80^\circ$C. 
 d)PS-PEO diblock with $f_A=0.5$ and $N_{\rm PS-PEO}=86$ and $N_{\rm H,PS}=16$ with $6\%$ LiFTSI salt doping, \cite{TwoSolvBD_Bates}[Figure 8a].
 e) From $\GDF$ model with $\chi N_{AB}=34.4$ other parameters as in Figure\,\ref{f:BD-LamA1}. 
  f) Ternay blend of PS-PEO diblock and PEO/PS homopolymers with $f_{\rm PEO}=0.4$, $N_{AB}=77$, $N_{\rm PEO}=6$, $N_{\rm PS}=10$, equal homopolymer volume fractions, and $6\%$ LiFTSI salt doping, from \cite{Bates-Lodge16}[Figure 2] 
 }
 \end{tabular} 
\end{center}

\begin{figure}[h!]
   \centering
\includegraphics[height=1.35in]{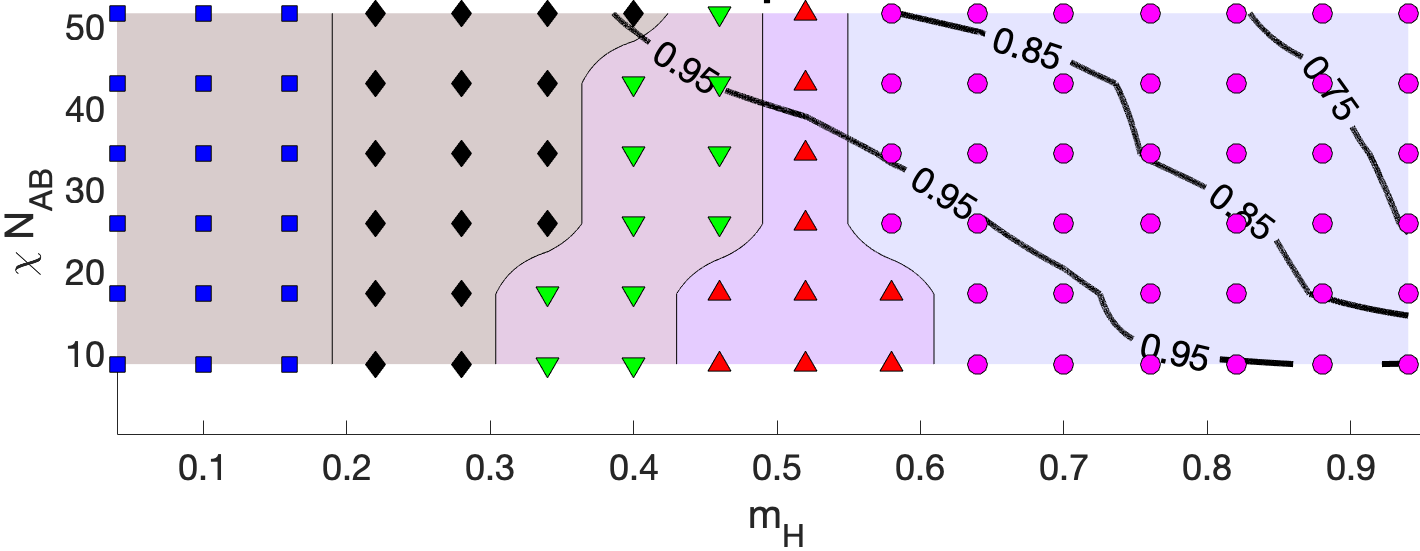} 
     \hspace{0.1in}
     \includegraphics[width=2.95in, height=1.35in]
     {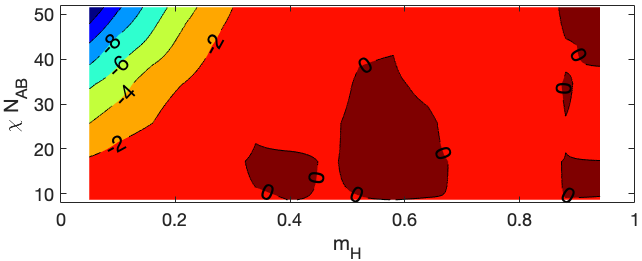}
      \put(-490, 90){\Large\textbf{(a)}}   \put(-226, 90){\Large\textbf{(b)}} 
    \caption{(a)   Dominant morphology in the AB-A binary blend and contour lines of $\mrI_{\rm AH}$ as function of homopolymer fraction $m_H$ and $\chi N_{AB}$ for perturbed uniform initial data with $f_A=0.5$ and $\epsA=1$. (b) Contour lines of differences in end-state energy, measured in k$_B$T/(100nm)$^3$ domain, between seeded and perturbed uniform initial data.
    }
    \label{f:BD-CST}
\end{figure}

The choice of initial data has nontrivial impact on the end-state equilibrium of the gradient flow. To quantify this the $m_H-\chi N_{AB}$ phase diagram was repeated replacing the seeded initial data of Figure\,\ref{f:BD-LamA1} with spatially uniform initial data subject to 10\% zero-mass white noise fluctuations in volume fraction. The results are reported in Figure\,\ref{f:BD-CST} (a). The two-phase regions demarcated by the  $\mrI_{\rm AH}=0.95$ level set are largely unchanged. The impact on microphase regions is significant, with the micellular region expanding at the expense of the hexagonal region. The pure-lamellar phases are replaced with a fluctuating lamellar phase with a relatively disordered interface.  Most significantly the region of lamellar-hexagonal coexistence (circled blue squares in Figure\,\ref{f:BD-LamA1}(a)) are replaced by a microemulsion type phase in Figure\,\ref{f:BD-CST}. These phases are evocative of the experimental results of \cite{Bates-Lodge16}[Figure 2] for PEO-PS ternary blends with  $N_{AB}=70$, $N_{HA}=10$, $N_{HB}=6$, and $f_A=0.5$ with lithium salt ions in a 6\% molar ratio to  ethylene oxide units. For equal volumes of homopolymer and total homopolymer mass fraction $m_H=m_{\rm H,PEO}+m_{\rm H,PS}$ in the range $[0.5 0.7]$ they report a transition from coexisting lamellar plus hexagon phases to microemulsion and microemulsion plus lamellar with increasing temperature (equivalently decreasing $\chi$). 
The ternary blends become two-phase at higher values of $m_H.$  The $\GDF$ gradient flow end-states depicted in Figures\,\ref{f:BD-LamA1}(a) and \ref{f:BD-CST}(a) show similar transitions. Indeed for $\chi N_{AB}=43$ the end state equilibrium from seeded initial data transition from lamellar to mixed lamellar-hexagonal at $m_H=0.25$ to two--phase at $m_H=0.35$ while for perturbation uniform initial data the mixed phases are replaced by a microemulsion phase and the transition to two-phase is delayed to $m_H=0.45$, see Table 1 (e, f). While the chain lengths in the $\GDF$ simulations and experiments are similar, direct comparison between a binary and ternary blend is indirect. The second homopolymer phase can stabilize the lamellar structures while the salt dopant has significant impact on the effective Flory-Huggins parameters. Nevertheless, there is strong visual agreement between the microemulsion phases computed from the $\GDF$ gradient flow, see Figure\,\ref{f:bistable} (b$^\prime$), and experimentally observed microemulsions \cite{Bates14}[Figure 4b, c] and \cite{Bates-97}[Figure 3b, c].

Spatial coexistence of microphase structures in the $\GDF$ gradient was only observed from seeded initial data.  All gradient flow end states obtained from the perturbed uniform initial data were comprised of a single microphase morphology. The differences in energy of the end states resulting from gradient descent from seeded and unseeded initial data are compared in Figure\,\ref{f:BD-CST} (b).  Despite variations in microstructure the energy differences, measured in k$_{\rm B}$T/(100 nm)$^3$, are small. The energies are roughly identical in the area below the diagonal of the Figure, and the energy of the perturbed uniform initial data exceeds that of the seeded data by at most $10$ k$_{\rm B}$T/(100 nm)$^3$ for the highest value of segregation and lowest value of $m_H$. The corresponding gradient flow end-states are presented in Figure\,\ref{f:bistable} (a-a$^\prime$). 
It is possible that very slow transients beyond those resolved by the simulations could allow microemulsion type phases to rearrange into a more ordered lamellar or hexagonal pattern or in coexisting lamellar and microphase morphologies observed in 
\cite{Bates14}[Figure 4d]. 

The SCMF continuation computations fail to predict bicontinuous microemulsion phases in ternary mixtures with $\alpha_A$ near 0.2 in which they are observed experimentally, \cite{Masten-BmuE20}. Instead they find two-phase mixtures of  pure homopolymer and lamella. Field theoretical models call into question the low temperature existence of microemulsions, instead suggesting a combination of two-phase mixtures of homopolymer with microemulsion transitioning to homopolymer plus lamella with reducing temperature (increasing segregation strength). 

\subsection{Structure of the Two Phase domains}

Transitions to two-phase structure have been tied to increases $\alpha_A$ the ratio of homopolymer to total diblock copolymer chain lengths. SCMF simulations suggest that the onset of two phase behavior is primarily dictated by the entropy of the chains, and that this drives larger chains to become immiscible and phase separate from the diblock copolymers.  The energy of the two-phase separation is reported to be large compared to the differences in energies between microphases, hence the onset of two phase behavior is relatively independent of the form of the microstructure within the copolymer rich phase \cite{Matsen95}.  The onset of two-phase behavior can also be triggered by homopolymer volume fraction at the unbinding point of the microstructure,  \cite{Schick-97}[Figure 1,2] and \cite{JanSch-97}[Figure 1]. The distribution of homopolymer within the microphase separated regions and the size and distribution of pure homopolymer domains after onset of two-phase behavior are of considerable interest to LCC filtration applications.  It is hypothesized that homopolymers will congregate in regions of high energy, in particular regions of high interfacial curvature.  It has been argued that an uneven distribution of a swelling agent can facilitate the formation of the C14 and C15 quasiperiodic micelle-phases \cite{ANS_21}.

\begin{figure}[h!]
   \centering
    \includegraphics[height=2.2in]{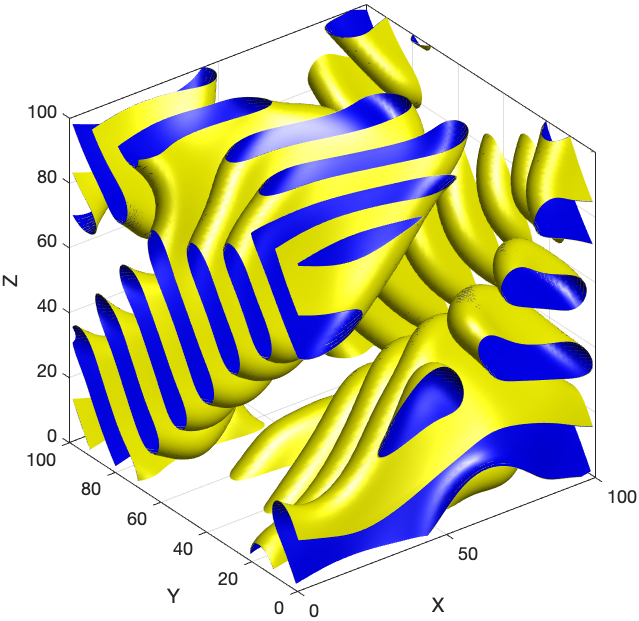}
    \includegraphics[height=2.0in]{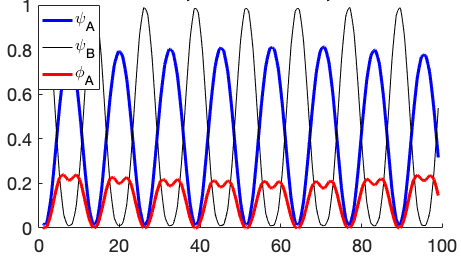}
     \put(-440, 130){\Large\textbf{(a)}}   \put(-265, 130){\Large\textbf{(d)}} \\
     \includegraphics[height=2.2in,width=2.2in]{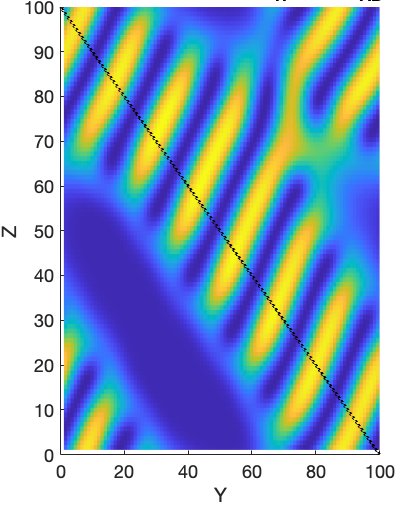}
     \hspace{0.05in}
        \includegraphics[height=2.2in,width=2.2in]{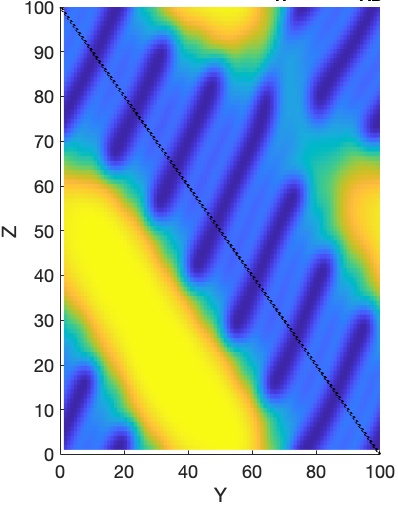} 
            \put(-340, 145){\Large\textbf{(b)}}   \put(-170, 145){\Large\textbf{(c)}} 
        \vspace{0.2in} \\
       
    \caption{ Microphase separation in a two-phase blend  with $f_A=0.5$,  $m_H=0.46, \chi N_{AB}=43$, 
$\alpha=.21$, $\epsA=1.67$. The AH projection satisfies $\mrI_{\rm AH}=0.445$. 
    (a) Level sets of volume fraction $\psi_A$ at 50\% and 52\% of maximum value reveal  lamellar microstructure intertwined with regions of pure $\psi_H$, depicted as void space. Cross section, at $X=40$nm of (b) $\psi_A$ diblock phase and (c) $\phi_H$ homopolymer. The yellow/blue colors denote regions of high/low volume fraction respectively. (d) The trace of the values of $\psi_A$ (thick blue), $\psi_B$ (thin black), and $\phi_H$ (thick red) taken along the diagonal dotted line (b, c) show the detailed bilayer structure  within the lamellar region. 
}
    \label{f:drybrush}
\end{figure}

The morphology of two-phase structures in the $\GDF$ energy is well represented by end-state corresponding to the gradient flow from seeded initial data with $\chi N_{AB}=43$ and $m_H=0.46$ in Figure\,\ref{f:BD-LamA1} (b). This simulation has $f_A=0.5$, $m_H=0.46, \chi=0.5$, $\alpha=0.21$, and $\epsA=1.67$ with projection $\mrI_{\rm AH}=0.45$, which is well inside the two-phase regime.  The 3D image in Figure\,\ref{f:drybrush}(a) shows the volume fraction of the $A$-block $\psi_A$ at level sets equal to 50\% and 52\% of its maximum value. There is a well-defined lamellar structure that is interspersed with homopolymer-rich regions (void). Figures\,\ref{f:drybrush}(c) and (d) show the $A$-block volume fraction, $\psi_A$, and homopolymer volume fraction $\phi_H$ respectively taken from the cross section of Figure\,\ref{f:drybrush}(a) in the $X=40$nm plane. The lamellar pattern shows swelling of the $A$-block by homopolymer, however the inclusions of almost pure homopolymer (bright yellow in Figure\,\ref{f:drybrush}(d)) containing the majority of the homopolymer volume. The $\GDF$ model with $\epsA=1$ has a moderate energetic cost to form two-phase domains, which facilitates ejection of homopolymer once swelling encounters significant energetic cost. The regularity of the lamellar pattern and degree of swelling is quantified in the 1D trace, taken along the dotted lines of Figure\,\ref{f:drybrush}(c) and (d) and shown in Figure\,\ref{f:drybrush} (b). The homopolymer and $A$-block have a 4-1 volume ratio within the lamellar region with a slight dip in homopolymer volume in the center of $A$-block domain.

\begin{figure}[h!]
   \centering
     \includegraphics[height=2.2in]{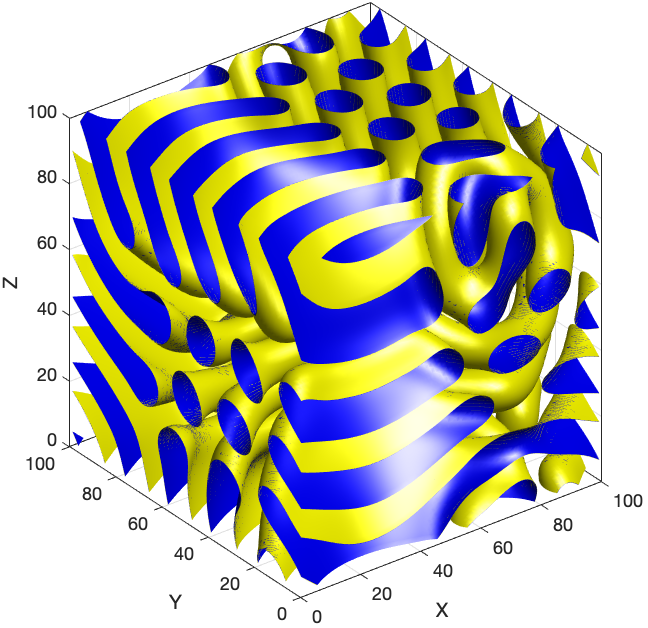}
      \includegraphics[height=2.0in]{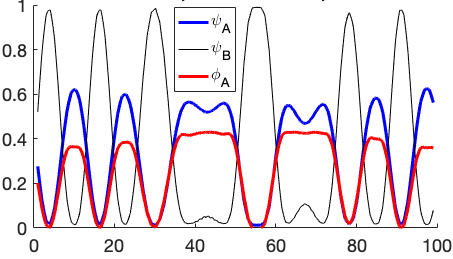}
          \put(-440, 130){\Large\textbf{(a)}}   \put(-265, 130){\Large\textbf{(d)}} \\
    \includegraphics[height=2.2in,width=2.2in]{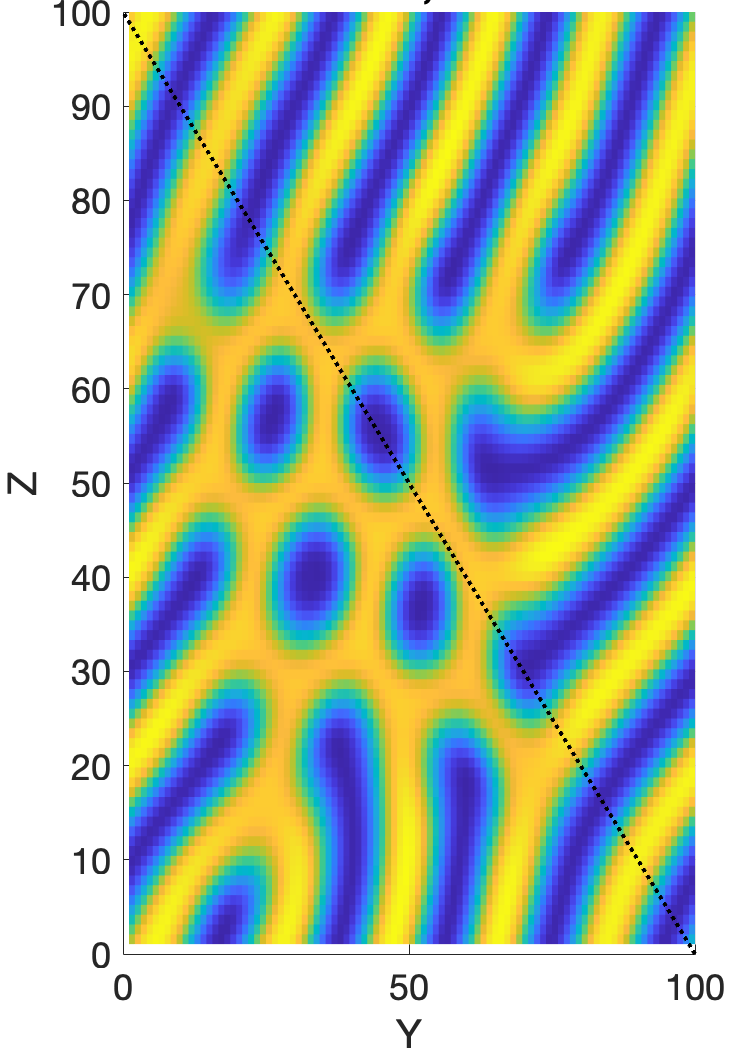}
         \hspace{0.08in}
        \includegraphics[height=2.2in, width=2.5in]{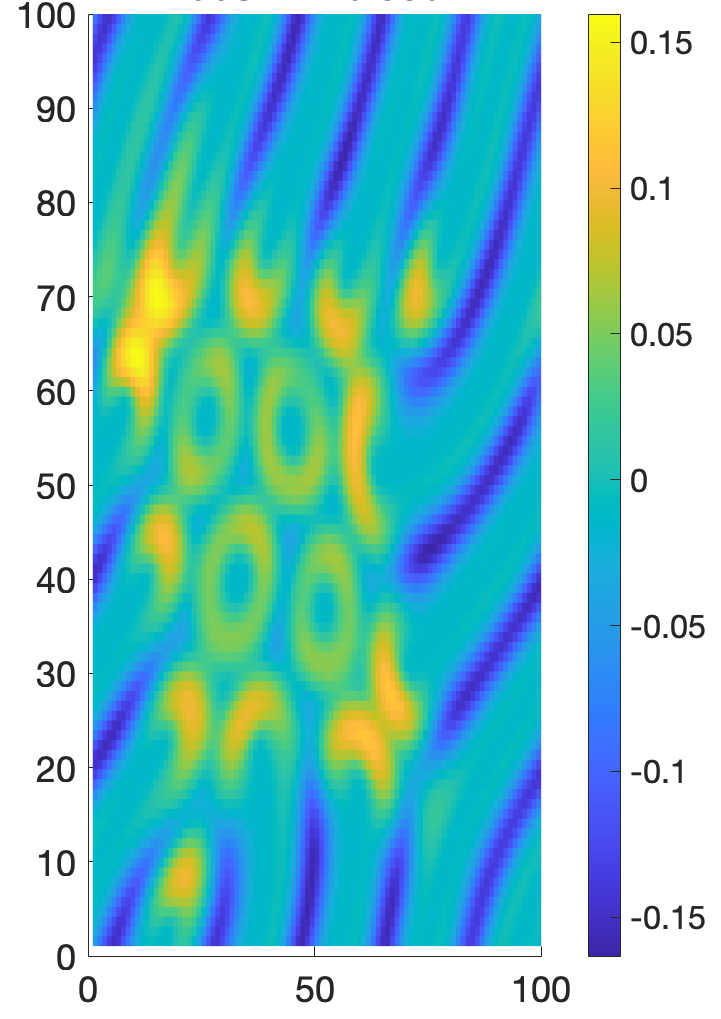}
    \put(-360, 145){\Large\textbf{(b)}}   \put(-197, 145){\Large\textbf{(c)}} 
       
    \caption{(a) Level sets depicting coexisting lamella and hexagonally packed cylinders generated from seeded initial data with $m_H=0.28, N_{AB}\chi=34.4$, $f_A=0.5,$ and $\epsA=1.0$.  The projection $\mrI_{\rm AH}=0.990$ corresponds to a wet brush regime.  Cross sections, at $X=40$nm show (b) the $\psi_A$ diblock density and  (c) the  homopolymer deviation $\delta_H$ \eqref{e:HPD}.  (d) The trace of $\psi_A$ (thick blue), $\psi_B$ (thin black), and $\phi_H$ (thick red) taken along the diagonal dotted line in (b) as function of $Y$. The lamellar structures ($Z\in[0,30]$ and  cylindrical pore $Z\in[50,70]$. The density of $\phi_H$ is proportional to $\psi_A$ in the lamella but overshoots at the edge of the cylindrical pore.}
    \label{f:HPdist}
\end{figure}
The $\GDF$ gradient flow end-state from seeded initial data corresponding to the $m_H=.28$, $\chi N_{AB}=34.4$, $f_A=0.5$, and $\epsA=1$ provides insight into the distribution of homopolymer within the microphase structure in the wet-brush regime.  From the $m_H-\chi N$ phase diagram of Figure\,\ref{f:BD-LamA1}(a) the corresponding end-state has coexisting lamellar and hexagonal domains, as confirmed by the level sets of $\psi_A$ presented in Figure\,\ref{f:HPdist}(a). With $\mrI_{\rm AH}=0.990$ the end-state is well within the wet-brush regime.  The cross-section taken along the $X=40$nm plane reveals details of the coexisting hexagonal and lamellar structures (Figure\,\ref{f:HPdist}(b) with a spatial arrangement that is influenced by the structure of the initial seeding. The homopolymer density is roughly proportional to the $A$-block volume fraction, however it has significant deviations. The homopolymer deviation function quantifies this departure of the homopolymer volume fraction from proportionally,
\beq\label{e:HPD}
\delta_{H}:=2\frac{m_{A}\phi_H-m_{H}\psi_A}{\phi_H+\psi_A}.
\eeq
This is shown in Figure\,\ref{f:HPdist}(c), showing that the homopolymer volume fraction is in excess by approximately $15\%$ around the high-curvature regions associated to the cylindrical phase and the end-caps of the lamella. 
 These are regions where the $A$-block chains are splayed by the high curvature induced by the $B$-block core. The homopolymer lowers the energy by back-filling these areas.
The traces of $\psi_A, \psi_B$, and $\phi_H$ along the dotted diagonal of the $X=40$nm plane of Figure\,\ref{f:HPdist}(b) are shown in Figure\,\ref{f:HPdist}(d). The variation in the ratio of $\phi_H/\psi_A$ is clearly dependent upon morphology. The local maxima of $\psi_A$ are highest in the relatively narrow regions within the $B$-block lamellar structures, and are lower in the longer defect regions between the cylindrical cores and the lamellar endcaps.

\begin{figure}
    \centering
     \includegraphics[width=5.2in]{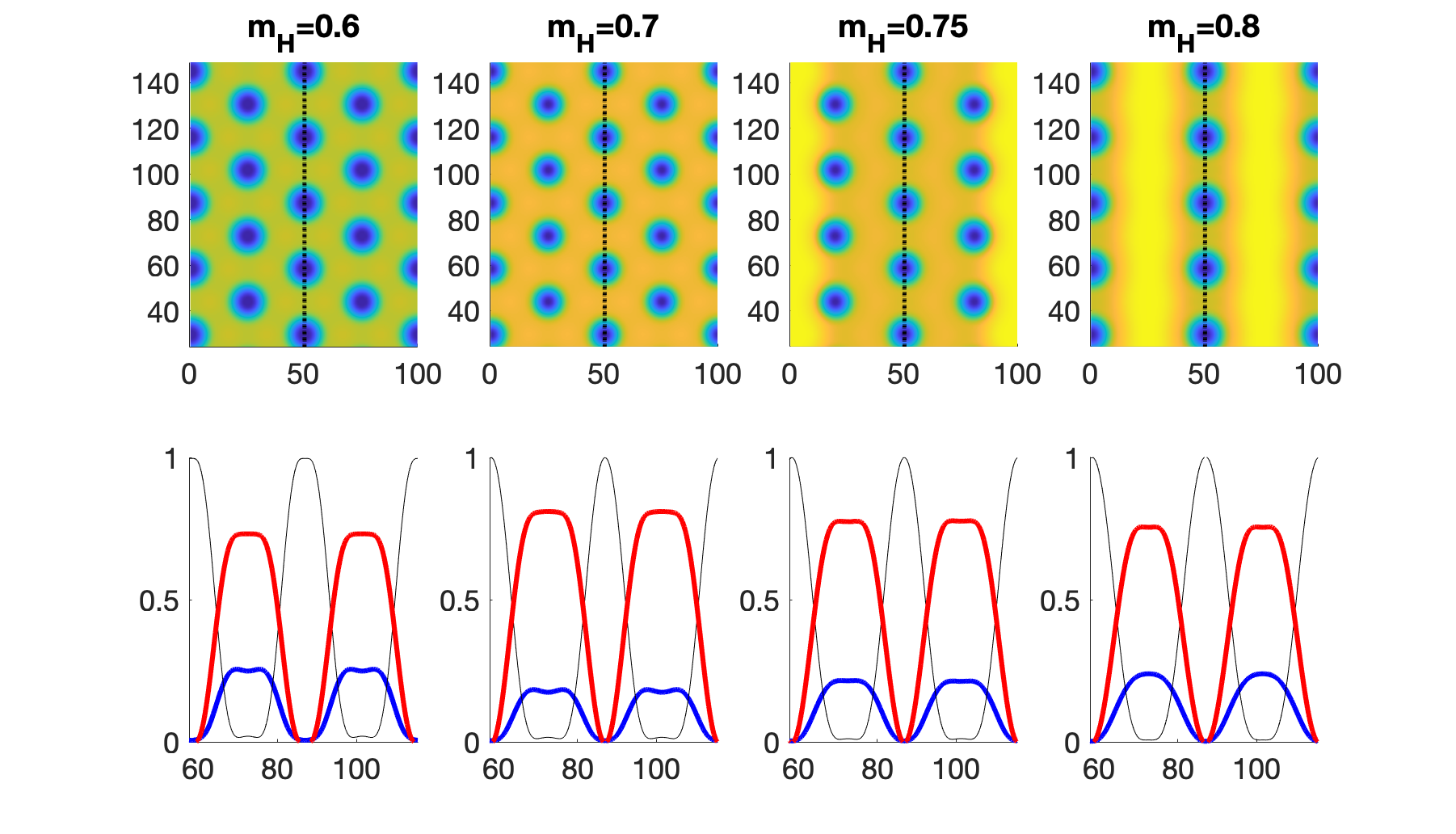}
\put(-350, 205){\Large\textbf{(a)}} 
\put(-350, 105) {\Large\textbf{(b)}}
     \caption{(a) The volume fraction of homopolymer, $\phi_H$, for four cylindrical microphases excised from a larger  $[0,200]\times[0,173]\times[0,40]$ nanometer-scaled domain. Parameter values are $f_A=0.5$, $\epsA=1.0, \chi N_{AB}=34.4$, and $m_H= 0.6, 0.7, 0.75, 0.8$. (b) The traces of $\psi_A$ (blue), $\psi_B$ (black-thin), and $\phi_H$ (red) taken over the range $Y\in[60,120]$nm of the dotted lines in the corresponding figure above.}
     \label{f:HEX}
\end{figure}

The lattice of hexagonally packed cylinders is known to swell under changes in homopolymer density, and forms a benchmark in the study of binary and ternary blends. To examine this in the $\GDF$ energy an initial condition of defect-free hexagonally packed cylinders is formed with $f_A=0.5$, $\epsA=1.0$, $N_{AB}\chi=34.4$ for $m_H=0.6$, and continued numerically for $m_H\in[0.6,0.8]$. These parameter choices cover the range of hexagonal cylinders in the $\chi N_{AB}=34.4$ row of Figure\,\ref{f:BD-LamA1}(a).  The hexagonally packed cylinders corresponding to the end state of $m_H=0.6$ are used to populate an equilibrium on a $[0,200]\times[0,173]\times[0,40]$ nanometer-scaled domain whose $x-y$ plane supports an exact hexagonal lattice. The value of $m_H$ in small increments to allow the system to reequilibrate.  Figure\,\ref{f:HEX} shows the structure of the homopolymer volume fraction for $m_H=0.6, 0.7, 0.75, 0.8.$ and traces of all three volume fractions along a subset of the $X=55$nm line.
For $m_H<0.75$ increasing the homopolymer volume fraction while maintaining $f_A=0.5$ shrinks the core of the hydrophobic $B$-block, increasing the volume allocated to the blend of homopolymer and $A$-block blend, while diluting the density of the $A$-block.  For $m_H$ beyond $0.7$ the volume of hydrophobic block becomes too small to support the number of cores determined by the hexagonal packing and the system rearranges into a two-phase blend with regions of almost pure $\phi_H$ and regions of hexagonally packed cylinders. Further increases in $m_H$ towards $0.8$ induces a reconfiguration into a cubic array of cylinders interspersed with pure homopolymer regimes. As can be seen from the traces taken along the symmetry line through the cylinder cores, see Figure\,\ref{f:HEX}(b), the maximum value of $\phi_H$ along the symmetry line increases to a critical value of approximately $0.8$ for $m_H=0.7$. Further increases in $m_H$ do not increase this value, rather excess homopolymer is ejected into the pure homopolymer regions. The trace of $A$-block volume fraction initially decrease with increasing $m_H$ but relax back towards the profile at $m_H=0.6$ after the two-phase transition.  The projection $\mrI_{\rm AH}$ remains  above $0.99$ for the symmetric, defect-free microstructures, but transitions to $I_{AH}=0.83$ and $0.75$ after the two-phase onset at $m_H=0.75$ and $0.8$ respectively. These values of $\mrI_{\rm AH}$ are comparable to those from both seeded and random initial data.
     
\begin{figure}
    \centering
    \includegraphics[height=2.0in]{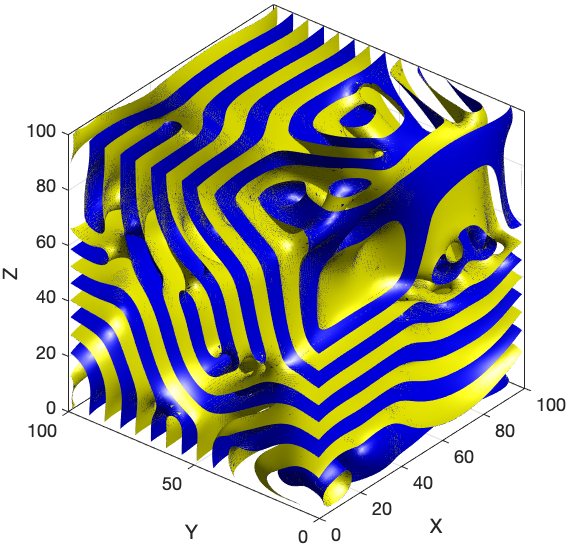}
     \includegraphics[height=2.0in]{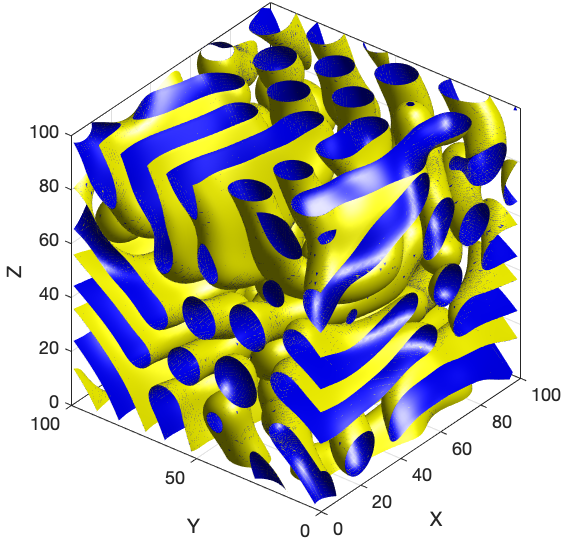}
      \includegraphics[height=2.0in]{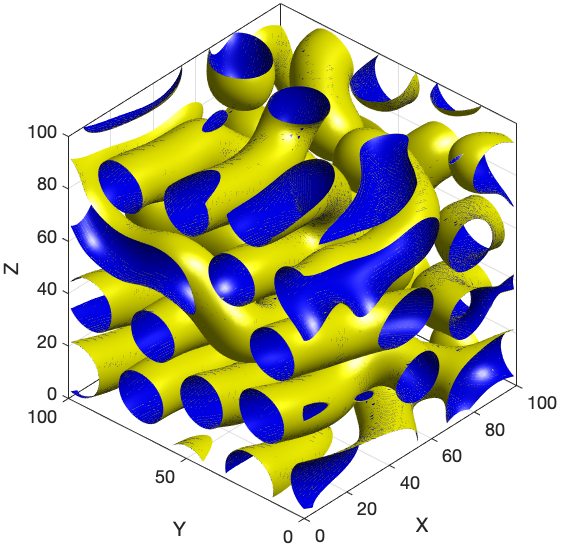}
          \put(-450, 130){\Large\textbf{(a)}}
          \put(-300, 130){\Large\textbf{(b)}}
     \put(-150, 130){\Large\textbf{(c)}} 
      \\
      \includegraphics[height=2.0in]{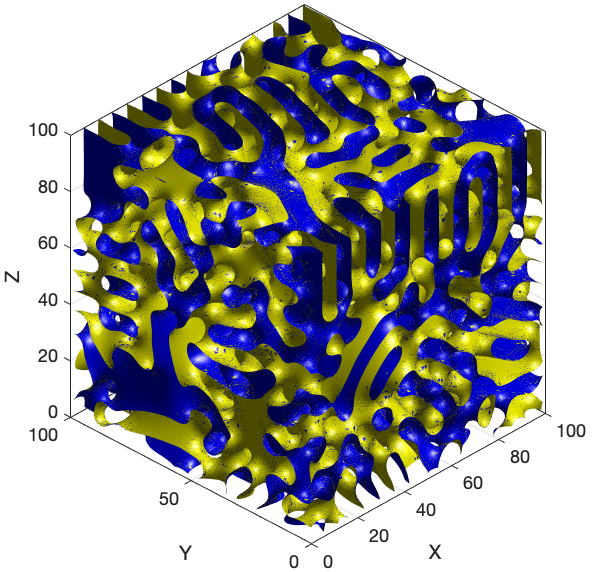}
     \includegraphics[height=2.0in]{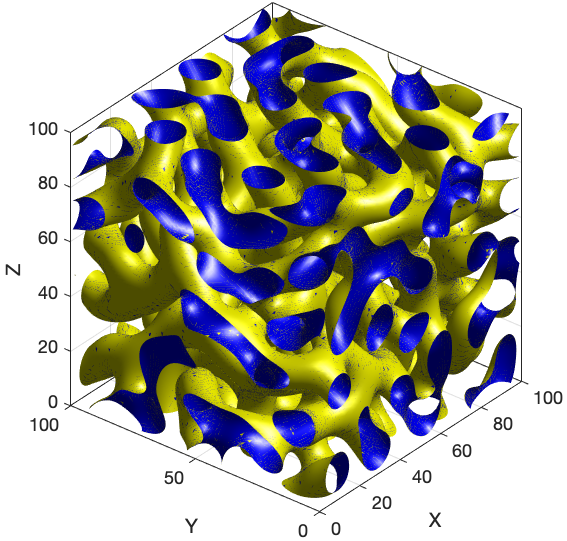}
      \includegraphics[height=2.0in]{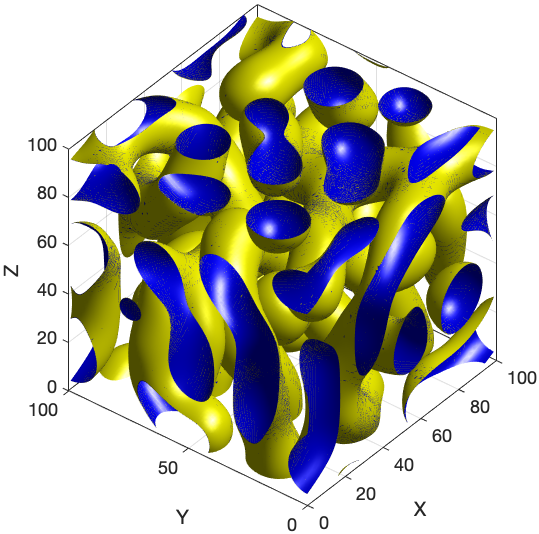} 
       \put(-450, 130){\Large\textbf{(a$^\prime$)}}
          \put(-300, 130){\Large\textbf{(b$^\prime$)}}
     \put(-150, 130){\Large\textbf{(c$^\prime$)}}
 
    \caption{Comparison of end-state of $\GDF$ gradient descent with seeded (unprimed) and perturbed uniform (primed) initial data. All simulations have  $\epsA=1$, and are (a) lamellar and (a$^\prime$) fluctuated lamellar with $\chi N_{AB}=43, m_H=0.04$, (b) mixed lamellar-hexagonal and (b$^\prime$) microemulsion with $\chi N_{AB}= 34.4, m_H=0.34$ and (c) hexagonal and (c$^\prime$) hexagonal with $ \chi N_{AB}=17.2, m_H= 0.46$. }
    \label{f:bistable}
\end{figure}

 \subsection{Numerical SAX data and Bistablity}
SAX data is fundamental to the investigation of in-situ morphology of microphase separated blends. For simulations, notwithstanding the presence of visualizations, SAX can assist in the classification of non-symmetric phases as it extracts weight from structural features that are not immediately apparent upon visual inspection.  The public domain code of Rohr \cite{Rohr-07}, is used to simulate SAX data for the end-state microphase morphologies. In particular SAXS data provides an algorithmic technique to identify the physical coexistence of multiple microphase domains through the presence of a small $q$ upturn in scattering amplitude.
For a numerical scattering intensity $I$, the empirical condition 
\beq
\label{e:Small-q}
I(q_0)> \frac12 \max_{q>0} I(q),
\eeq
is effective at detecting coexisting microphase domains in the wet-brush regime. Here $q_0$ is the smallest value of scattering parameter $q$, corresponding to a length scale of roughly half the domain size.   

\begin{figure}
    \includegraphics[width=3.3in] {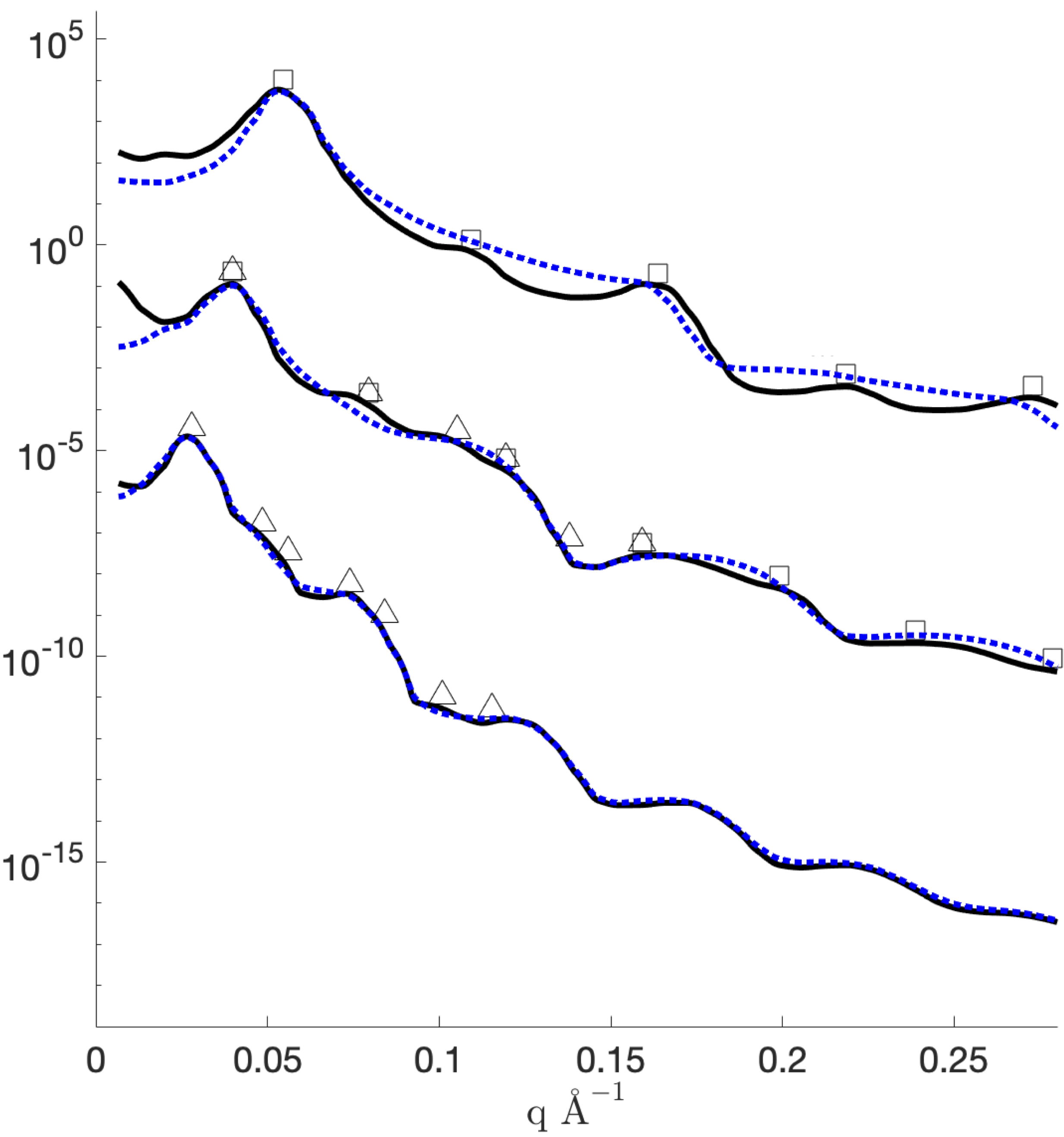}  
      \put(-35, 170){\textbf{a-a$^\prime$}}
       \put(-35, 115){\textbf{b-b$^\prime$}}
        \put(-35, 60){\textbf{c-c$^\prime$}}
    \caption{Semilog plot of scaled scattering intensity corresponding to the three pairs of morphologies from seeded (solid line) and perturbed uniform  (dotted line) initial data shown in Figure\,\ref{f:bistable}. Lamellar and hexagonal peak multiples are indicated with squares and triangles respectively.}
    \label{f:SAX}
    
\end{figure}


The study of seeded verses perturbed uniform initial data revealed the onset of a microemulsion type phase as indicated in Figure\,\ref{f:BD-CST}(a) that was not present in the gradient descent from seeded initial conditions. The differences in pairs of end-states for seeded verses perturbed uniform initial configurations are presented in Figure\,\ref{f:bistable} for three of homopolymer volume fraction and segregation strength. The first pair (a-a$^\prime$), with $m_H=0.04$ and $\chi N_{AB}=43$ shows lamellar verses fluctuated lamellar microphases, the second pair (b-b$^\prime$) with $m_H=0.34$ and $\chi N_{AB}=34.4$  depicts coexisting lamellar and cylindrical domains verses a microemulsion morphology, and the third pair (c-c$^\prime$) with with $m_H=0.46$ and $\chi N_{AB}=17.2$  presents hexagonally packed cylinders verses a more disordered hexagonal packing. The energy difference between the end-states is largest for (a-a$^\prime$) pair and roughly vanishes for the (c-c$^\prime$) pair.
The SAX scattering data  presented in Figure\,\ref{f:SAX}
illuminates the differences between these $\GDF$ end state pairs. They are roughly equivalent for material parameters for which end-state pairs had comparable energies, such as (c-c$^\prime$).  For end-states with larger energy differences  the SAX intensities diverge. Comparing the lamellar and fluctuated lamellar microphases, the scattering intensity for the lamellar phase displays the classical SAX structure with peaks at integral multiples $\{1, 2, 3, 4, 5, \ldots\}$ of the primary peak.
The fluctuating lamellar phase has a similar primary peak however its secondary peaks are suppressed, and there is no small $q$ upturn, indicating an absence of longer range spatial correlation, see Figure\,\ref{f:SAX}(a-a$^\prime$). 
For the (b-b$^\prime$) pair, the SAX data for the coexisting lamellar and hexagonal domains shows peaks at both the lamellar and the hexagonal multiples $\{1, \sqrt{3}, \sqrt{4}, \sqrt{7}, \sqrt{9}, \sqrt{13}, \ldots\}$ of the primary peak. There is a significant small-$q$ upturn, associated to long-range correlations induced by the competing microphase domains. The end state from perturbed uniform initial data, Figure\,\ref{f:bistable}(b$^\prime$), has no significant secondary peaks and no small-$q$ upturn due to its spatial homogeneity on longer length scales, see Figure\,\ref{f:SAX}(b-b$^\prime$).  Small $q$ upturns are frequently but not universally observable in experimental SAX, \cite{Bates-Lodge16}[Figure 8], \cite{TwoSolvBD_Bates}[Figure 7] and may be impacted by auxiliary processing.
For the third pair the end states from seeded and perturbed uniform initial data are both cylindrically dominated phases, Figure\,\ref{f:bistable} (c-c$^\prime$). The SAX intensities are nearly identical with secondary peaks at the hexagonal multiples, see Figure\,\ref{f:SAX}(c-c$^\prime$).
The SAX curve agreement holds despite the lower symmetry of the end-state from the perturbed uniform initial data.

\begin{figure}
  \includegraphics[width=3.0in]{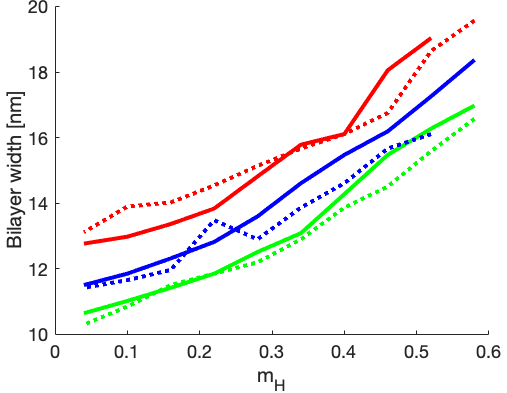}
 \caption{Lamellar width verses homopolymer volume fraction for $\epsA=1.0$ (solid) and $\epsA=1.67$ (dotted) for $\chi N_{AB}=34.4, 43, 51.6$ (red, blue, green).
  }
  \label{f:BLW}
\end{figure}
  
In lamellar phases the primary peak of the SAX intensity is associated to the lamellar spacing. In diblock copolymer-homopolymer blends the peak location can be used to characterize the rate of swelling of the lamellar phases with increased homopolymer volume fraction.  The lamellar width data is presented in Figure\,\ref{f:SAX}(b) as extracted from the SAX peaks associated to the end states presented in Figure\,\ref{f:BD-LamA1} with $\chi N_{AB}= 34.4, 43, 51.6$.  The lamellar width increases by roughly $50\%$ as the homopolymer fraction $m_H$ is increased from 0.04 to 0.6. Larger segregation strengths lead to thinner lamellar spacing. Smaller values of $\epsA$ generally lead to thinner lamellar spacing, in particular when the mixture has entered a two phase regime.  These results are comparable to experimentally observed lamellar swelling of 45\%-70\% over a comparable range of $m_H$ for a ternary blend \cite{TwoSolvBD_Bates}[Figure 5].

\section{Reduction to Scalar Microemulsion Models}

A melt of a binary AB-A blend of diblock co-polymer and homopolymer in a wet-brush regime can be approximately characterized by a single volume fraction.  For $\mrI_{\rm AH}$ close to unity, the density of homopolymer $\phi_{H}$ is close to a fixed multiple of the A-block density, $\psi_A$. In this case minimization of the free energy can be considered over classes of volume fractions satisfying $
\phi_H=\frac{m_{H}}{m_A} \psi_A.$ Combining this with the incompressibility constraint \eqref{e:incomp}, yields the relation
$$ \psi_B=1-\gamma \psi_A.$$
where $\gamma=(m_A+m_H)/m_A\geq 1$ indicates the volumetric swelling of $A$-diblock phase by the $A$ homopolymer. With this reduction the binary $\GDF$ model can be described in terms of a single variable $u:= \gamma \psi_A$  for which $\psi_B=1-u$, $\psi_A=u/\gamma$, and $\phi_H=(\gamma-1)/\gamma u.$ One additional modification is required to recover the microemulsion free-energy models proposed by Teubner and Strey, Gompper and Schick, and Gommper and Kraus. The Flory-Huggins interaction between the $\phi_B$ and $\phi_H$ densities must be replaced with an interaction between $\psi_B$ and $\phi_H$. After the $u$-substitution above, this replacement is equivalent to exchanging the interaction energy
$$\epsA\int_\Omega u \mrG^{-1} u=\epsA \int_\Omega |\mrG^{-\frac12} u|^2\geq 0,$$ 
with the local interaction term 
$$\int_\Omega |u|^2\geq 0.$$ 
Modulo this modification the $\GDF$ energy takes the microemulsion form 
\beq
\label{e:GG}
\cF_{\rm \mu EM}(u) = \int_\Omega \overline{a} |\nabla^2 u|^2 + g(u) |\nabla u|^2 + f(u) \, dx
\eeq
where the constant $\overline{a} = \epsA^2(1/(m_H\gamma^2)+1/m_B)/12$, and the functions $f$ and $g$ take the form
\beq
g(u) =\epsA^2\left( \frac{b_{AA}}{\gamma u} -\frac{2b_{AB}}{\sqrt{\gamma u(1-u)}} +\frac{b_{BB}}{1-u}\right)
        -\frac{2\epsA^2\chi}{\gamma} +\frac{\gamma^2}{12m_{H}},
\eeq
and
\beq
f(u) = \epsA^2\left(\frac {c_{AB}}{\gamma u} - \frac{2c_{AB}}{\sqrt{\gamma u(1-u)}}+ \frac{c_{BB}}{1-u}\right)(u-m_A-m_{H})^2 + \frac{1-\gamma}{\gamma}u^2+\frac{\gamma}{N_{H}}u \ln u.
\eeq
The coefficients $b_{ij}$ an $c_{ij}$  depend only upon the chain lengths and are given in the supplementary material. 
For $\chi$ sufficiently large the potential $g$ is positive for $u$ near $0$ and $1$, and negative for $u$ near $\frac12$, which meets the essential structural feature of the energy and associated scattering intensity found in \cite{TS-87}[eqn 4], \cite{GS-90}[eqn 1], and \cite{Gompper-K_I-93}[eqn 1] for strong amphiphiles. 

\section{Conclusions}

 The parameterization of the SSE embeds a large class of averaging transformations into the the RPA derivation of density functional models  while  preserving key structural features. The simple $\GDF$ model presented has a single fit parameter and is not tuned to fit any particular data set.  It supports phase diagrams that reproduce the full range of structures observed in complex polymer blends. It gradient flow shows sensitive dependence upon initial configurations, the coexistence of microphase domains, and bistability. It supports microemulsion phases. The simple transformation  $\cG=\mrG/\La$ connects the RPA to the scalar microemulsion scattering structure models of Tuebner and Strey \cite{TS-87}, Gompper and Schick, \cite{GS-90}, and Gompper and Kraus \cite{Gompper-K_I-93} that sought to characterize microemulsion phases. By adjusting the form of the transformation the model can interpolate between these microemulsion models and the traditional RPA models of Ohta-Kawasaki and their extensions by Uneyama and Doi. 
 
 The $\GDF$ model has shortcomings, chief among them is a blurring of the ODT line and a two-phase regime with an unbinding transition at a lower homopolymer mass fraction than the SCMF models. While the full parametric family of $\eDF$ models share the same scattering structure at the ODT point, the singular Fourier symbol in the $\GDF$ model strengthens the tendency of the model to generate interface and weakens the sharp dichotomy present in the Ohta-Kawasaki spinodal decomposition, replacing it with a transition that accommodates bistability.  Different choices of the averaging transformation, in particularly one that is closer to the identity, would restore the sharpness of the ODT.  
 This suggests that DF models will have to be sensitively tuned to be accurate across the huge state space presented by a complex polymer blend.
 
 The broad family of parameterized DF models is a strong candidate for machine-learning based force matching to finer scale DPD, SCMF, and field theoretical models. They are capable of reproducing the full range of features obeserved in experiments and simulations of finer-grained models.  The averaging transformations embedded in the parameterized SSE step are natural proxies for the mid- to long-range order induced by polydispersity, salt, ionic groups, and domain formation. Ideally the broad family of parameterized DF models can serve as accurate templates for finer scale models in patches of parameter space, and then be woven into a global surrogate model which can efficiently scan high dimensional parameter spaces. Such a surrogate model can be naturally coupled to non-equilibrium auxiliary processes that are essential for control of the microstructure that optimizes filtration and conduction properties of membranes. 

\section{Supplementary Material}

\subsection{Overview of RPA reduction of mean field energy}
A critical point analysis of the free energy derived in the self-consistent mean field yields a system for the entropy of the diblock chains. These are governed  by a Fokker-Planck equation for the forward and backward propagators $q$ and $q^*$ for the polymers.  Associated to an AB diblock chain of length $N_{AB}$ the backward propagator satisfies
$$ q_s + (l^2/6) \Delta q - U_{AB} q =0,$$
subject to $q(y,N_{AB})=1$ and $(y,s)\in \Omega\times (0,N_{AB}).$ The force is constant in $s$ on each fraction of the diblock
$$U_{AB}(x,s)=\left\{\begin{array}{ll} U_A(x), & s\in[0,f_AN_{AB}], \\
U_B(x), & s\in[f_AN_{AB},N_{AB}],
\end{array}\right.$$ 
where $f_A$ is the $A$ diblock fraction. The forward propagator satisfies 
$$q^*_s - (l^2/6) \Delta q^* + U_{AB} q^* =0,$$
subject to $q^*(y,0)=1$. Here $s$ denotes length along the polymer chain and $l$ is the Kuhn length describing the 
stiffness of the polymer chain. The volume fraction is recovered from the relations
$$ \phi_i(x) = \frac{n_i}{\int_\Omega q(y,0)\,dy} \int_{\mathcal{I}_i} q(x,s)q^*(x,s)\,ds,$$
where $n_i$ is the number of polymer chains of type $i=A,B$ and $\mathcal{I}_i$ is the collection of all polymer chains composed of monomer type $i$.  

 A key step within the RPA is the long-chain approximation to invert the relation between the volume fractions and external force at a linear level. These relations are block diagonal over the molecular classes. For an AB diblock with components $i,j\in\{A,B\}$, this leads to the relation between the volumetric deviation $\phiro_i$ and the external forces
\beq
\label{e:hR-def}
\widehat{\phiro_{i}}(\zeta) = -  \sum_{j=i,i'} \widehat{R}_{ij}(\zeta)\widehat U_{j}(\zeta),\eeq
in the frequency space.
Here 
$R$ is a $2\times2$ symmetric operator defined through
\beq
\widehat{R}_{ij}=\left\{ \begin{aligned}
&2m_i \frac{h(cf_iN_{AB})}{c^2}, & j=i,\\
&m_j\frac{g(cf_iN_{AB}, cf_jN_{AB})}{c^2}, & j=i',
\end{aligned}\right.
\eeq
where $m_i$ is the volume fraction of the diblock molecule of type $i$, and $N_{AB}$ is the chain length of the $AB$ diblock copolymer.  This quantity is expressed in terms of the modified Debye functions
\begin{align}
\label{e:c-def}
c(\zeta)&:=\ell^2 |\zeta|^2\big/6, \\
 h(s)&:=e^{-s}+s-1, \nonumber\\
 g(s_1, s_2)&:=\left(1-e^{-s_1}\right)\left(1-e^{-s_2}\right).
 \nonumber
\end{align}
Inverting $R$ yields the  volume fraction to external force relation
\beq
 \label{e:hT-def} 
 \widehat U_i(\zeta) = - \sum_{j=i,i'} \widehat{T}_{ij}(\zeta) \widehat {\phiro_{j}}(\zeta).
 \eeq 
These are approximated using  the long and short wave-length expansions of $h$ and $g$: 
\begin{align}
 h(s)\sim s, &\qquad  g(s_1, s_2) \sim 1, &\textrm{for}\quad  cf_AN_{AB}\gg 1, cf_{B}N_{AB}\gg 1,\\
 h(s)\sim s^2/2 -s^3/6,\qquad & g(s_1, s_2)\sim (s_1-s_1^2/2)(s_2-s_2^2/2), &\textrm{for}\quad cf_AN_{AB}\ll 1, cf_{B}N_{AB}\ll 1.
\end{align}
The approximations follow Ohta-Kawasaki, but include the cubic term in the small $s$ expansion of $h$ to preserve invertability. 
 In the physical space, for each diblock copolymer of length $N_{AB}$, the operator $T=(T_{ij})$ admits an expansion in terms of powers of the Laplacian $\Delta$,
\beq
\label{e:Feynman-Kac}
T_{ij}= -\frac{\ell^2 }{m_j} \Delta T_{ij}^{(1)} +\frac{1}{N_{AB}} \frac{1}{\sqrt{m_if_im_{j}f_{j}}} T_{ij}^{(0)} +\frac{1}{N_{AB}^2\ell^2} \frac{1}{\sqrt{m_if_im_{j}f_{j}} } T_{ij}^{(-1)}(-\Delta)^{-1}.
\eeq   
The $2\times2$ matrices $T^{(n)}$ depend only upon the diblock fractions,
\beq
\label{e:T-ell}
\begin{aligned}
&T^{(1)} =\frac{1}{12}\left(\begin{array}{cc}
1 & 0\\
0 & 1
\end{array}\right);\quad 
T^{(-1)}= 9\left(\begin{array}{cc}
1 & -1\\
-1 & 1
\end{array}\right) +\frac{3}{4}\left(\begin{array}{cc}
f_{A}^{-1} & 0\\
0 & f_{B}^{-1}
\end{array}\right);\\
& T^{(0)}= 
\frac{1}{16} \left(\begin{array}{cc}
9-8f_{A}-22 f_{A}f_{B} & 17 -22f_{A}f_{B} \\
17-22f_{A}f_{B} & 9-8f_{B}-22 f_{A}f_{B}
\end{array}\right). 
\end{aligned}
\eeq
Within the binary energy the copolymer matrices $T^{(n)}$
are rewritten in  terms of the interaction matrices,
\beq\label{eq:adef1}
a_{ij} = T_{ij}^{(1)}, \quad
b_{ij} = \frac{T_{ij}^{(0)}}{N_{AB} \sqrt{f_if_{j}}}, \quad 
c_{ij} = \frac{T_{ij}^{(-1)}}{N_{AB}^2\sqrt{f_if_{j}}}.
\eeq
For a homopolymer of type $i$, length $N_i$, and total volume fraction $m_i,$ the corresponding operator $T=T_i$ is scalar valued and takes the form
\beq
\label{e:T-HP}
T_i = -\frac{\ell^2}{12m_i}\Delta +\frac{1}{m_i N_i}.
\eeq 
For a solvent of type $i$ with chain length $N_i=1$, and total volume $m_i$ the operator $T=T_i= \frac{1}{m_i}.$

\subsection{Binary Diblock/Homopolymer melt}

The simplest $\GDF$ model describes a binary blend  homopolymer of monomer $A$ with an $A-B$ diblock. In terms of $\psi=(\psi_A,\psi_B,\phi_H)$ the energy has the explicit form
\beq\label{eq:AB-H}
 \begin{aligned}
 \cF_{\rm 
 \GDF}(\psi) =& \int_\Omega  \epsA^2\bigg[
      \frac{|\nabla^2 \psi_A|^2} {12m_A} +   \frac{|\nabla^2\psi_B|^2} {12m_B} + 
      \frac{b_{AA}|\nabla\psi_A|^2}{\psi_A}  +\left( \frac{2b_{AB}}{\sqrt{\psi_A\psi_B}} +2\chi\right)\nabla\psi_A\cdot\nabla\psi_B\\
      & \hspace{0.3in}
      +\frac{b_{BB}|\nabla \psi_B|^2}{\psi_B} +
      \frac{c_{AA}(\psi_A-\bphi_A)^2}{\psi_A} +2\frac{c_{AB}(\psi_A-\bphi_A)(\psi_{B}-\bphi_{B})}{\sqrt{\psi_A\psi_{B}}} 
\\
&\hspace{0.2in} +\frac{c_{BB}(\psi_B-\bphi_B)^2}{\psi_B}\biggr] +  2\epsA\chi\phi_H \mrG^{-1}\psi_B +
\frac{|\nabla \phi_H|^2}{12m_{H}} +  
\frac{\phi_H\ln\phi_H}{N_{H}} 
\,\textrm{d}x.
\end{aligned}
\eeq
For $i,j\in\{A,B\}$, the coefficients $b_{ij}$ and $c_{ij}$  from \eqref{eq:adef1} depend solely upon the diblock chain structure,
\begin{align*}
(b_{ij}) & =
\renewcommand\arraystretch{2}
\begin{bmatrix}
\dfrac{9-8f_A-22 f_Af_B}{16N_{AB} f_A} & \dfrac{17-22 f_Af_B}{16N_{AB} \sqrt{f_A f_B}}\\
\dfrac{17-22 f_Af_B}{16N_{AB} \sqrt{f_A f_B}}  & \dfrac{9-8f_B-22 f_Af_B}{16N_{AB} f_B}\\
\end{bmatrix}, \\
(c_{ij}) & =
\renewcommand\arraystretch{2.5}
\begin{bmatrix}
\dfrac{\frac{3}{4}+9f_A}{N_{AB}^2 f_A^2} & \dfrac{-9}{N_{AB}^2 \sqrt{f_A f_B}} \\
\dfrac{-9}{N_{AB}^2 \sqrt{f_A f_B}}  & \dfrac{\frac{3}{4}+9f_B}{N_{AB}^2f_B^2} \\
\end{bmatrix}. 
\end{align*}
The point-wise incompressibility constraint
\beq 
\phi_H = 1 -\psi_{A}- \psi_B,
\eeq
reduces the free energy to a function of $(\psi_A,\psi_B)$ only.

\subsection {Ternary Diblock/Homopolymer melt}
For a  blend of an AB-diblock with A and B type homopolymers the ternary $\GDF$ free energy is written in terms of $\psi=(\psi_A,\psi_B,\phi_{HA},\phi_{HB})$
\beq\label{eq:ternary}
 \begin{aligned}
 \cF_{\rm \GDF}(\psi) = & \int_\Omega \Biggl( \epsA^2\bigg[
      \frac{|\nabla^2 \psi_A|^2} {12m_A} +   \frac{|\nabla^2\psi_B|^2} {12m_B} + 
      \frac{b_{AA}|\nabla\psi_A|^2}{\psi_A}  +\left( \frac{2b_{AB}}{\sqrt{\psi_A\psi_B}} +2\chi\right)\nabla\psi_A\cdot\nabla\psi_B +\frac{b_{BB}|\nabla \psi_B|^2}{\psi_B} + \\
      & \hspace{0.3in}      +
      \frac{c_{AA}(\psi_A-\bphi_A)^2}{\psi_A} +2\frac{c_{AB}(\psi_A-\bphi_A)(\psi_{B}-\bphi_{B})}{\sqrt{\psi_A\psi_{B}}} +
\frac{c_{BB}(\psi_B-\bphi_B)^2}{\psi_B}\biggr] + \\
&\hspace{0.2in} 2\epsA\chi\biggl[ \phi_{HB}\mrG^{-1}\psi_A + \phi_{HA} \mrG^{-1}\psi_B\biggr] +
\frac{|\nabla \phi_{HA}|^2}{12m_{HA}} + 
\frac{|\nabla \phi_{HB}|^2}{12m_{HB}} + \\
&\hspace{0.2in}
\frac{\phi_{HA}\ln\phi_{HA}}{N_{HA}}+
\frac{\phi_{HB}\ln\phi_{HB}}{N_{HB}}+
2\chi \phi_{HA}\phi_{HB}\Biggr) \,\textrm{d}x.
\end{aligned}
\eeq
This is subject to the incompressibility condition $\psi_A+\psi_B+\phi_{HA}+\phi_{HB}=1.$
\subsection{Mollification of Singularity}
To develop an efficient numerical scheme for the gradient computations the second order implicit-explicit (IMEX) approach proposed in \cite{FCH-Benchmark} was employed. This requires mollification of the singularities present in the $\GDF$ energy. The singular terms are extended so that the energy is large but finite in the regions $\psi_i<0$ and $\psi_i>1$. This is accomplished by developing 4th order Taylor polynomial interpolants for key singular terms. In particular  $\ln{x}$ is replaced with 
$$f(x)=T(x)\chi_{[\infty,\delta]}+\ln(x)~\chi_{[\delta,\infty]},$$ 
where $\chi$ is the characteristic function of the interval, and $T(x)$ is a fourth order Taylor polynomial of the natural logarithm at $x=\delta\ll1$. The value of $\delta=10^{-3}$ is taken so that incompressibility conditions $\psi_i\in[0,1]$ are imposed to within $10^{-3}.$ Other singular functions, such as $1/x$ and $1/\sqrt{x}$ are approximated by $f^\prime$ and $\sqrt{f^\prime}$ to preserve relations between derivatives of these functions within the structure of the energy and gradient flow.



\bibliographystyle{unsrt}
\bibliography{MicroemulsionDFT}

\end{document}